\documentclass[9pt,shortpaper,twoside,web]{ieeecolor}
\usepackage{subfiles}

\usepackage{generic}
\usepackage{cite}
\usepackage{amsmath,amssymb,amsfonts}
\usepackage{algorithmic}
\usepackage{graphicx}
\usepackage{multirow}
\usepackage{textcomp}
\usepackage{hyperref}
\usepackage{capt-of}
\usepackage{supertabular}
\usepackage{multibib}
\newcites{supp}{Supplementary References}
\usepackage{lipsum}

\def\BibTeX{{\rm B\kern-.05em{\sc i\kern-.025em b}\kern-.08em
    T\kern-.1667em\lower.7ex\hbox{E}\kern-.125emX}}
\begin{document}

\title{Multimodal Data Integration for Precision Oncology: \\ Challenges and Future Directions}
\author{Huajun~Zhou, \IEEEmembership{Member,~IEEE,} Fengtao~Zhou, Chenyu Zhao, Yingxue~Xu, Luyang Luo, \IEEEmembership{Member,~IEEE,} Hao~Chen*, \IEEEmembership{Senior Member,~IEEE.}
\thanks{This work was supported by the Hong Kong Innovation and Technology Fund (Project No. MHP/002/22) and Research Grants Council of the Hong Kong Special Administrative Region, China (Project No. R6003-22 and C4024-22GF). (Corresponding author: Hao Chen.)}
\thanks{Huajun Zhou, Fengtao Zhou, Chenyu Zhao, Yingxue Xu, and Luyang Luo are with the Department of Computer Science and Engineering, The Hong Kong University of Science and Technology, Hong Kong, China.}
\thanks{Hao Chen is with the Department of Computer Science and Engineering, Department of Chemical and Biological Engineering and Division of Life Science, Hong Kong University of Science and Technology, Hong Kong, China. (e-mail: jhc@cse.ust.hk). }
}
\maketitle

\begin{abstract}
The essence of precision oncology lies in its commitment to tailor targeted treatments and care measures to each patient based on the individual characteristics of the tumor.
The inherent heterogeneity of tumors necessitates gathering information from diverse data sources to provide valuable insights from various perspectives, fostering a holistic comprehension of the tumor.
Over the past decade, multimodal data integration technology for precision oncology has made significant strides, showcasing remarkable progress in understanding the intricate details within heterogeneous data modalities. 
These strides have exhibited tremendous potential for improving clinical decision-making and model interpretation, contributing to the advancement of cancer care and treatment.
Given the rapid progress that has been achieved, we provide a comprehensive overview of about 300 papers detailing cutting-edge multimodal data integration techniques in precision oncology. 
In addition, we conclude the primary clinical applications that have reaped significant benefits, including early assessment, diagnosis, prognosis, and biomarker discovery.
Finally, derived from the findings of this survey, we present an in-depth analysis that explores the pivotal challenges and reveals essential pathways for future research in the field of multimodal data integration for precision oncology.

\end{abstract}


\section{Introduction}
\label{sec:introduction}
According to the estimation provided by the International Agency for Research on Cancer (IARC) of the World Health Organization (WHO), we witnessed 20 million new cases of cancer and, unfortunately, 9.7 million cancer-related deaths in 2022 \cite{WHOdata}. 
Cancer patients usually have a high mortality rate within five years of cancer diagnosis and endure significant mental, financial, and physical burdens.
In addition to being an important barrier to increasing life expectancy, cancer is associated with substantial societal and macroeconomic costs that vary in degree across cancer types, geography, and gender \cite{chen2023estimates}.
Precision oncology represents a pivotal paradigm in cancer treatment, aiming to tailor therapeutic approaches based on the distinctive characteristics of patients' tumors. 
By customizing treatment plans to maximize efficacy while mitigating adverse effects, precision oncology holds immense promise in improving treatment outcomes and advancing the landscape of cancer care.
Nevertheless, the intricacies inherent in the micro- and macro-environment of tumors, coupled with the diverse characteristics exhibited by different cancers, present a significant challenge in comprehending the complex nature of tumors and devising more effective therapies.

\begin{figure*}[!t]
\centering
\includegraphics[width=1 \textwidth]{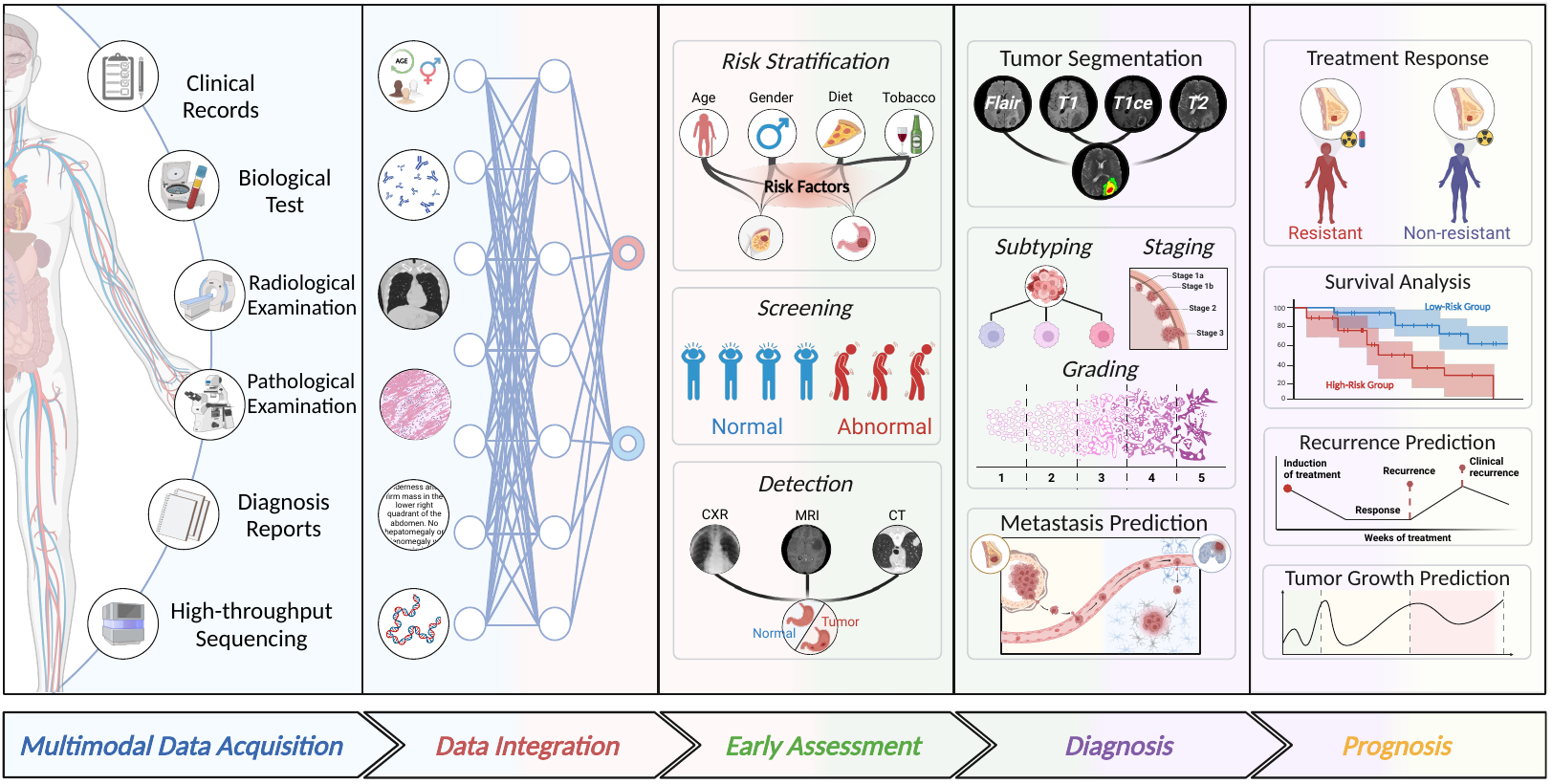}
\caption{Overview of multimodal data integration for advancing precision oncology.  }
\label{fig:motivation}
\end{figure*}

Clinicians have long relied on medical imaging \cite{bar2003clinical, makaju2018lung, morrow2011mri, tang2009computer} or lab test results \cite{chernecky2012laboratory, sturgeon2008national} to gain critical insights into patients' health conditions, enabling accurate diagnoses and informed treatment decisions.
In recent years, the precision oncology community has witnessed a surge \cite{zhuang20223d, zhou2023cross, xu2023multimodal, xu2021mufasa} due to the successful integration of a variety of heterogeneous data like medical imaging, clinical records, and omics data by leveraging multimodal data integration techniques.
Specifically, medical imaging provides detailed visualizations of the internal structures and abnormalities to enable the characterization of tumors, and assessment of their size, location, and spread.
Moreover, clinical records provide comprehensive insights into the patient's past and present health status, diagnostic findings, treatment approaches, and disease progression. 
Furthermore, omics data provides a deeper understanding of the molecular alterations associated with cancer, including the identification of genetic mutations, gene expression patterns, protein modifications, \textit{etc}. 
These heterogeneous data modalities provide valuable yet distinct insights into tumor characteristics, risk assessment, cancer progression, and treatment response. 
Effectively integrating these multimodal data offers promising opportunities for building a holistic understanding of tumors and advancing healthcare research, diagnostics, and personalized medicine, as shown in Fig. \ref{fig:motivation}.

However, constructing Artificial Intelligence (AI) models for effectively integrating multimodal data is a non-trivial task, requiring multi-faceted considerations on various critical aspects. 
These include understanding multimodal data characteristics, devising effective model architectures, formulating robust fusion strategies, and addressing potential challenges.
Specifically, for samples with complete modalities, the primary objective is to effectively integrate heterogeneous knowledge in different modalities to improve the model's performance. 
In the realm of multimodal data integration, there exist diverse fusion strategies possessing distinct advantages and drawbacks, calling for a thoughtful evaluation of the specific data modalities and clinical tasks to determine the most suitable fusion strategy.
Moreover, for samples with incomplete modalities, the focus shifts toward learning robust representations to minimize performance degradation.
Imputation-based methods focus on compensating the missing modalities using information from the observed modalities, while imputation-free methods directly leverage the observed modalities to perform multimodal fusion without imputing the missing modalities.
As the former method may introduce additional noise by imputing missing modalities, and the latter method overlooks the correlations between modalities, striking a balance between noise reduction and capturing inter-modality relationships becomes crucial.

\begin{figure}[!t]
\centering
\includegraphics[width=0.47 \textwidth]{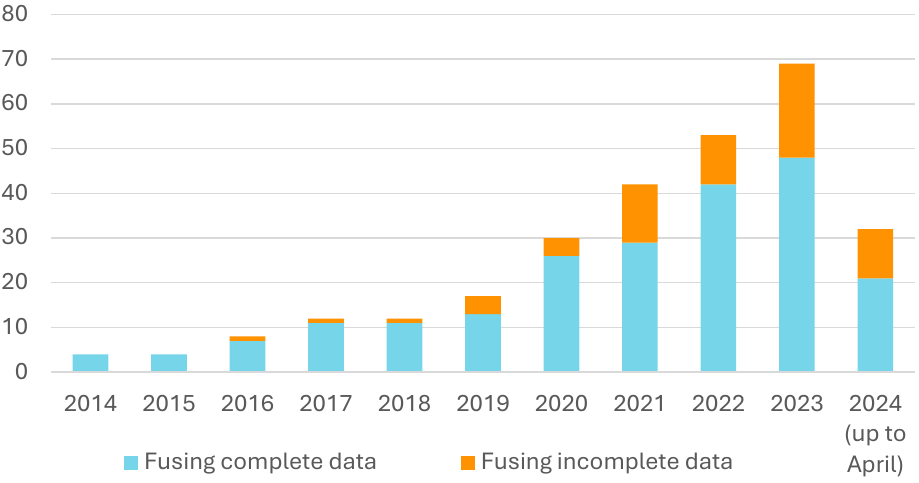}
\caption{Histogram of the reviewed papers on multimodal data integration for precision oncology in the past decade.}
\label{fig:histogram}
\end{figure}

In this paper, we surveyed about 300 publications in the field of multimodal data integration for precision oncology over the past 10 years (2014 - 2024 up to April), as listed in Fig. \ref{fig:histogram}.
This review stands out from existing literature \cite{kline2022multimodal, panayides2020ai, qiu2023large, tong2023integrating, kang2022roadmap, steyaert2023multimodal, baltruvsaitis2018multimodal, acosta2022multimodal, huang2020fusion, lipkova2022artificial, lahat2015multimodal, zhang2020advances, stahlschmidt2022multimodal, jacobs2023artificial, muhammad2021comprehensive, zhao2024review, kho2022saline, soomro2022image, boehm2022harnessing} on the specified research topic due to its extensive analysis of the strengths and limitations of methodologies utilized in clinical applications of precision oncology.
Specifically, we first categorize the reviewed methods into two main topics mainly based on their different focus in dealing with complete or incomplete data.
For samples with complete modalities, we further categorize existing methods into early, intermediate, late, and multi-level fusion, and subsequently conduct an in-depth analysis of their properties regarding architectural complexity, multimodal interconnection modeling, and the potential risk of modality collapse.
These critical aspects are instrumental in ensuring optimal effectiveness and efficiency in leveraging multimodal data integration for precision oncology applications.
For samples with incomplete modalities, we provide a detailed categorization of imputation-based methods, specifically into three distinct subcategories: data generation, feature generation, and sample retrieval, and imputation-free methods, specifically into three distinct subcategories as well: robustness enhancement, multi-task learning, and knowledge distillation.
This refined categorization allows for a more comprehensive understanding and exploration of the various approaches employed to address the challenges posed by incomplete modalities.
Furthermore, our investigation delves into the clinical applications of multimodal data integration within the context of precision oncology. 
By exploring the practical use cases, we discuss the challenges that impede the advancement of multimodal data integration in the realm of precision oncology. 
By identifying these challenges, we explore potential future directions for further advancements in integrating diverse data modalities to enhance precision oncology approaches and improve patient outcomes.

The remainder of this work is structured as follows: 
In Section \ref{sec:modality}, we illustrate data modalities and corresponding modality representation extraction techniques.
Next, in Section \ref{sec:method}, we review existing multimodal data integration techniques from two perspectives, complete and incomplete data, respectively.
Subsequently, we investigate the clinical applications of multimodal data integration in Section \ref{sec:application}.
Based on the above investigation, we conclude several challenges and potential future directions in Section \ref{sec:challenge}.
Finally, we summarize our survey in Section \ref{sec:conclusion}.

\section{Data Modality}
\label{sec:modality}

\textbf{Imaging data} provides valuable visual information that helps clinicians in diagnosing cancers, assessing the extent and progression of conditions, planning treatments, and monitoring treatment responses.

Endoscopic and dermoscopic images are captured using seamlessly integrated cameras within their respective instruments, namely endoscopes and dermatoscopes. 
These cameras utilize diverse imaging techniques, such as white-light and narrowband images, that are considered distinct modalities.
Given the similarity to natural images, existing encoders pre-trained on natural images, such as CNN \cite{ronneberger2015u,çiçek20163d} or Transformer \cite{dosovitskiy2020image, wang2021transbts} models, can be employed to extract deep feature from each image directly.

Radiology imaging technologies aim to show the structure or function of tissues and organs and are widely used for diagnosing and treating cancers. 
There are two main types of imaging: structural imaging, which creates images of the anatomy and morphology of body parts, and functional imaging, which captures the functioning of tissues and organs \cite{histed2012review}.
Structural imaging techniques include computed tomography (CT), magnetic resonance imaging (MRI), ultrasound (US), and mammography scans. 
CT imaging uses X-rays to create detailed cross-sectional images of the body, while MRI imaging employs a strong magnetic field and radio waves for detailed cross-sectional images. 
US imaging uses sound waves to generate real-time images of internal organs, and mammography uses low-dose X-rays for detailed images of breast tissue, making it the standard for breast cancer screening \cite{luo2024deep}.
Functional imaging techniques like positron emission tomography (PET), single-photon emission computed tomography (SPECT), and optical imaging reveal the functioning of tissues and organs.
SPECT and PET use small amounts of radioactive tracers to produce concentrated images of body parts, while optical imaging uses digital cameras to detect molecular emissions from electromagnetic waves. 
Molecular imaging targets specific biomolecules involved in cellular processes underlying disease states.
Despite the inherent differences between radiology images and natural images, existing encoders are commonly utilized for feature extraction.

Pathology image diagnosis represents the gold standard in tumor diagnosis, offering a meticulous examination of tissue structures and cellular characteristics, unrivaled by radiology scans. 
However, the high-resolution nature of pathology images poses a challenge for AI models to extract discriminative features while disregarding non-informative regions.
To leverage the rich information within pathology images, Multiple Instance Learning (MIL) \cite{ilse2018attention} has emerged as a prominent approach. 
Specifically, each pathology image is split into numerous image patches, while patch features are aggregated to form a holistic representation.
MIL strategy enables AI models to select informative patches and extract lower-dimension yet discriminative representations from pathology images.

\textbf{Clinical data} encompass a wealth of medical records from cancer patients, including medical history, medications, demographics, laboratory test values, diagnostic reports, \textit{etc.}
Structured data in clinical records refers to information organized in a predefined format, which may be continuous (\textit{e.g.}, age and tumor size) or discrete (\textit{e.g.}, race and metastasis status) variables.
To integrate them into a joint representation, various techniques are employed for continuous and discrete variables, respectively.
Continuous variables can undergo normalization to ensure comparability across scales. 
Meanwhile, discrete variables with limited categories can be transformed using one-hot encoding, where each category is converted into a binary feature. 
By aggregating all encoded features, structured data can be transformed into a cohesive representation, facilitating subsequent multimodal integration.
On the other hand, unstructured data in clinical records refers to information that is not organized in a predetermined format, such as free-text clinical notes and diagnostic reports.
They often require natural language processing (NLP) techniques to extract relevant information for subsequent analysis and decision-making.
It is noteworthy that structured data can also be formulated as sentences, allowing for a more comprehensive and nuanced understanding of clinical records.
Recent approaches leverage the Large Language Models (LLMs) \cite{thirunavukarasu2023large, xu2021mufasa} to capture complex semantic information in textual data, facilitating advanced comprehension of clinical reports.

\textbf{Omics data} refers to large-scale biological data generated from high-throughput technologies that capture information about various biological molecules, such as genes, proteins, and metabolites, which are considered different modalities \cite{zheng2023multi, han2022multimodal}.  
It finds extensive application in systems biology and functional genomics, enabling the exploration of molecular interactions and their impact on the overall functionality of cells, tissues, and organisms.
Omic data, characterized by complexity, high dimensionality, and noise, necessitates the utilization of specialized computational methods and tools for its analysis and interpretation.
To this end, researchers employ advanced techniques such as self-normalizing neural networks (SNN) \cite{klambauer2017self} to enable a deeper understanding of the underlying biological mechanisms, facilitating personalized medicine and targeted therapies.

\begin{figure*}[t]
\centering
\includegraphics[width=0.99 \textwidth]{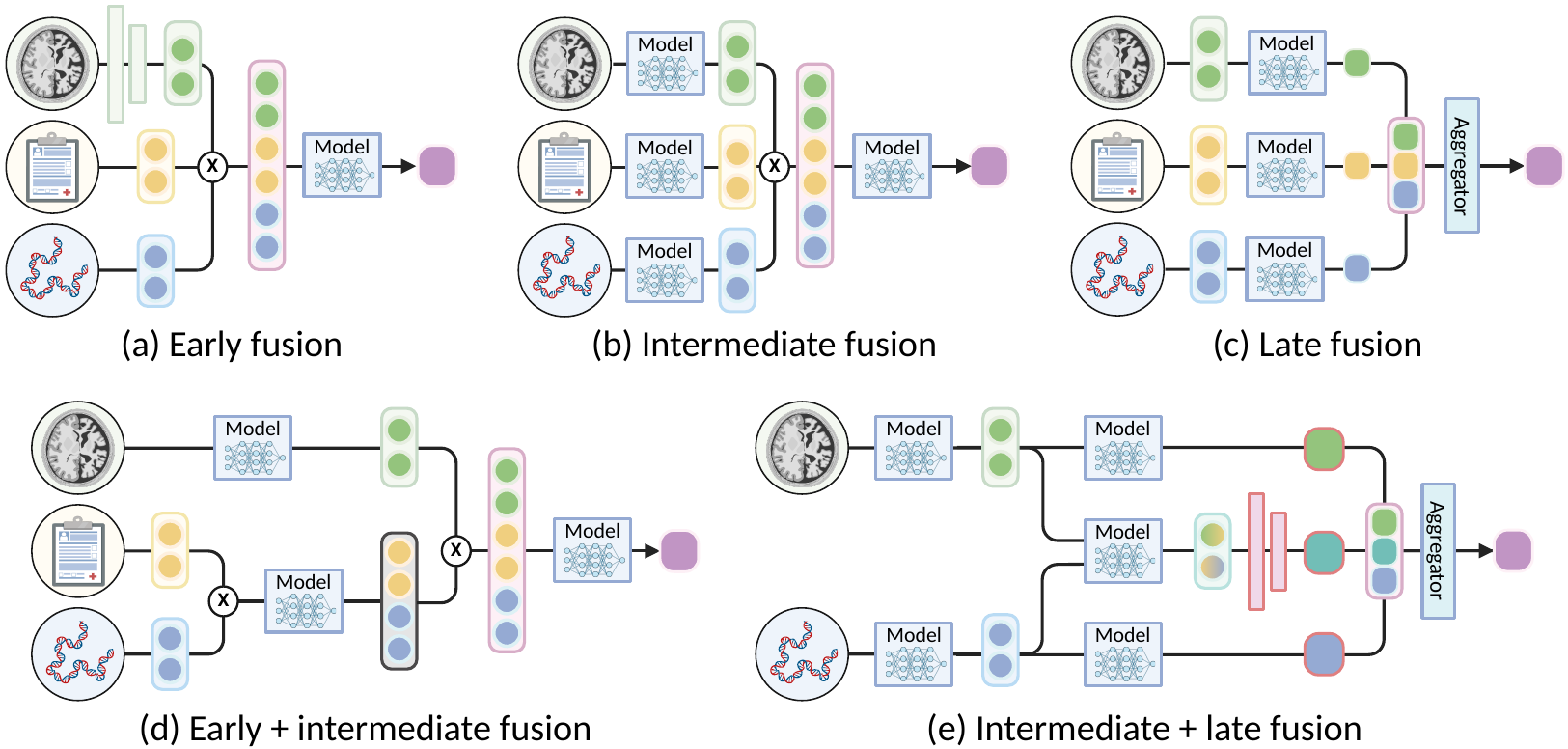}
\caption{Fusion strategies for complete data, including (a) early fusion, (b) intermediate fusion, (c) late fusion, and (d-e) multi-level fusion.}
\label{fig:taxonomic}
\end{figure*}

\section{Methods of Multimodal Data Integration}
\label{sec:method}
Multimodal data integration in precision oncology aims at leveraging heterogeneous information from multiple data sources to build a holistic understanding of tumors. 
When constructing AI models for multimodal data integration, two scenarios arise: samples with full modality data or some modalities are missing. 
Each scenario entails specific objectives for model construction, requiring careful consideration and adaptation based on data availability.
In the case of complete data, researchers strive to maximize the performance of downstream tasks by effectively integrating multimodal data.
Conversely, in incomplete cases, robust methods are necessary to handle incomplete data and minimize potential performance degradation.
Both scenarios offer unique opportunities to unveil patterns, enhance predictive accuracy, and facilitate informed decision-making in precision oncology.

\subsection{Integration of Complete Data}
To integrate multimodal data, we conclude four fusion strategies from the reviewed papers in Fig. \ref{fig:taxonomic}, including early, intermediate, late, and multi-level fusion.

\subsubsection{Early Fusion}
Early fusion refers to the integration of multimodal information at the input level, which could be raw data, hand-crafted features, or pre-processed deep features.
Concatenation is the most straightforward operation to obtain a joint representation \cite{xu2016multimodal, chai2021integrating, saeed2022tmss, fang2021self, gu2023segcofusion}, as it is capable of accommodating any format of representation.
Moreover, element-wise operations such as addition, multiplication, concatenation, or pooling can be adopted for modalities of the same shape, especially aligned multimodal imaging data.
For example, pixel-wise concatenation of different MRI sequences has been widely adopted in recent works \cite{tang2020deep, cheng2022fully, razzak2018efficient, yang2023flexible, hou2023mfd, qian2021prospective}.

Deep models designed for early fusion generally exhibit a relatively lower architectural complexity compared to other fusion strategies that involve processing multiple inputs simultaneously.
For instance, simple U-shape networks \cite{luo2020hdc, cui2020unified} can effectively extract joint representation from the concatenated multimodal inputs, as they operate on a single input stream.
The low architectural complexity of early fusion facilitates model design, parameter tuning, and interpretation, making them more accessible and convenient for clinicians.
While early attempts \cite{pereira2016brain, schulte2021integration, anagnostou2020multimodal} widely embrace it, the approach of early fusion has gradually faced some criticism and been overshadowed by more intricate fusion strategies in the latest works.

The first issue is the limitation on bridging explicit and intricate interconnections between multiple modalities.
Firstly, modalities often have different data types, structures, and scales \cite{toney2014neural}. 
Directly concatenating them into a unified input may make it difficult to effectively bridge intricate yet meaningful interconnections. 
The multimodal heterogeneity necessitates careful consideration of pre-processing steps and feature engineering techniques to appropriately integrate multimodal data.
Secondly, when concatenating multiple modalities, the dimensionality of the input substantially increases \cite{cui2020unified, tang2020deep}, leading to the accumulation of information redundancy across all modalities.
It presents a formidable challenge when dealing with high-dimensional inputs, as it demands a substantial amount of data to mitigate overfitting and effectively learn intricate patterns. 
The scarcity of available data, combined with the soaring dimensionality, gives rise to a sparsity predicament, impeding the ability to achieve good generalization on unseen samples.
Thirdly, the interaction between different modalities can manifest in intricate and non-linear dynamics, introducing a layer of complexity \cite{stahlschmidt2022multimodal} that may not be well captured by early fusion.
Certain multimodal interconnections may necessitate the utilization of specific attention mechanisms, gating mechanisms, or fusion techniques. 
Overlooking these interconnections will limit the model's capacity to fully leverage the heterogeneous information in multimodal data.

Another potential issue is modality collapse, wherein the learned representation excessively relies on a single modality while underutilizing information from other modalities \cite{wang2020auto}. 
Specifically, the primary goal of multimodal data integration is to effectively integrate information from multiple modalities, leveraging their complementary nature. 
However, modality collapse poses a significant challenge to this objective by limiting the contribution of certain modalities, leading to an imbalanced or biased representation. 
Consequently, the utilization of multimodal information is compromised, impeding the attainment of a comprehensive and accurate understanding of the data.
Within the early fusion approach, modality collapse can occur when one modality dominates the fusion process due to stronger predictive signals or when there exists a significant dimensionality gap between the modalities.
This imbalance can hinder the model's ability to capture the synergistic effects and complementary nature of different modalities, resulting in the underutilization of available multimodal information and suboptimal performance \cite{javaloy2022mitigating, nazabal2020handling}.

\subsubsection{Intermediate Fusion}
Intermediate fusion involves the fusion of multimodal information at the feature level, culminating in the extraction of an abstract joint representation for decision-making.
Given the diverse nature of feature modeling across different modalities, the fusion operations used in intermediate fusion exhibit significant variability. 
In addition to the concatenation operations employed in early fusion, intermediate fusion offers a wide range of additional operations that can be utilized, such as Graph Neural Networks (GNNs) \cite{liu2024muse, mo2020multimodal}, Transformers \cite{nakhli2023sparse, zhou2023cross, 10155265}, and attention mechanisms \cite{li2023survival, zhang2017tandemnet}. 
These techniques provide more flexibility in capturing the complex multimodal interconnections and enhancing multimodal representation.
Due to its inherent flexibility, intermediate fusion has garnered growing attention in recent works within the field of precision oncology. 

Intermediate fusion generally has a moderate architectural complexity. 
In the intermediate fusion approach, deep models showcase a sophisticated architecture that incorporates modality-specific sub-networks to capture the distinctive characteristics of each modality, along with fusion modules that model the interconnections between different modalities.
This additional processing step increases architectural complexity and computational requirements compared to early fusion, posing challenges like scalability concerns, computational efficiency, and the need for efficient training schemes.

The intricate interplay between sub-networks and fusion modules enables the capture of explicit and intricate interconnections between multiple modalities.
Existing approaches leverage multimodal data to enhance model performance from various perspectives.
The first approach is to align the feature representations of different modalities \cite{nakhli2023sparse, chen2019robust, ding2023pathology, ning2021multi} to emphasize consensus and improve the confidence of the predictions.
By mapping the features into a shared representation space, different modalities can be effectively inter-connected, enabling the exchange of information and enhancing consensus \cite{meng2024nama, lara2020multimodal, chen2021multimodal}.
Another approach involves leveraging the strengths of each modality and combining their complementary information \cite{zhang2017tandemnet, 8911262, fang2021self, gu2023segcofusion, zhang2024prototypical}. 
Rather than consensus enhancement, these methods seek to capture the unique contributions \cite{fu2021multimodal} of different modalities, enhancing the overall understanding and decision-making process.
Furthermore, some researchers aim to model the dependencies and interactions between modalities explicitly \cite{liu2022multimodal, liu2024multimodal, meng2022msmfn}. 
Graph-based representations, for example, enable the creation of a structure that reflects the interconnections between modalities \cite{mo2020multimodal, shi2024mif}, facilitating the propagation of information and capturing complex interactions.
The above approaches highlight the distinct merits of multimodal learning, focusing on common, unique, and synergistic knowledge between modalities, respectively.
To avoid favoring one type of knowledge over others and potentially overlooking valuable knowledge, comprehensive knowledge decomposition \cite{10155265, 9478224} has gained increasing attention. 
This approach involves decomposing multimodal knowledge into distinct components, allowing for a comprehensive analysis of each knowledge component's contributions. 
By incorporating all knowledge components and dynamically adapting their contributions, a holistic and nuanced comprehension can be attained, consequently yielding remarkable performance enhancements.
Besides, a quantification analysis \cite{liang2023quantifying} of knowledge components is crucial for evaluating the significance of each knowledge component, identifying potential biases or imbalances, and fine-tuning the model to ensure a fair and effective integration of all knowledge types.
However, this direction is still relatively underexplored, highlighting the necessity for further research and development.

Modality collapse presents a notable concern within intermediate fusion, as it occurs when the fusion process inadequately harnesses the information from all modalities, consequently leading to suboptimal performance. 
This issue can manifest in different ways, highlighting the need for careful consideration in multimodal fusion operations.
One common manifestation is the dominance of some modalities in the fusion process, overshadowing the contributions of other modalities. 
This scenario arises when one modality is more informative or easier to converge, causing the fusion model to heavily rely on that modality while neglecting the valuable information from other modalities.
Another manifestation occurs when the fusion process fails to effectively capture the complementary information in different modalities. 
Consequently, redundant or irrelevant information may be preserved in the fusion process, limiting the potential benefits of multimodal fusion.
To mitigate modality collapse, several techniques \cite{huang2022modality, wu2022scaling} have been explored to ensure a balanced integration of modalities, promote the equitable utilization of information, and capture the synergies between different modalities. 

\subsubsection{Late Fusion}
Late fusion aggregates modality-specific decisions into a more accurate joint decision, leveraging the decisive information from different modalities.
The aggregation operations in late fusion \cite{yala2019deep} may be the same as the concatenation operations used in early fusion.
In some cases, alternative aggregation operations, such as weighting \cite{fang2022weighted}, feature selection \cite{shao2019integrative,boehm2022multimodal}, rule-based aggregation \cite{yan2024combiner}, Bayesian-based fusion \cite{liu2023m} or learnable modules \cite{zheng2023multi}, can further enhance the fusion process.
These options provide a certain level of flexibility to customize the fusion process according to the specific characteristics of multimodal data.
In the reviewed papers, a considerable number of studies adopted this strategy.

In deep models, late fusion approaches typically exhibit a higher architectural complexity when compared to early and intermediate fusion \cite{zhao2024review, zhuang20223d}.
Specifically, the overall architecture of late fusion typically comprises multiple parallel branches, with each branch dedicated to a specific modality. 
This architecture allows for the incorporation of separate model structures that are tailored to capture the unique and nuanced characteristics of each modality. 
However, the architectural complexity arises from the need for processing multiple branches and ensuring proper aggregation of the modality-specific decisions. 
Overall, while late fusion approaches offer the advantage of capturing distinctive information from each modality, their added complexity may require additional computational resources and model parameters.

The utilization of modality-specific branches also introduces challenges in effectively leveraging the synergistic effects between different modalities. 
Specifically, when processing each modality independently, these individual branches may encounter limitations in capturing the complex interactions that arise when multiple modalities are combined \cite{zheng2023multi, fang2022weighted}.
These interactions play a crucial role in understanding the underlying interconnections between modalities and making more accurate joint decisions. 
Therefore, the absence of multimodal interactions hampers the ability to capture the complete knowledge and exploit the synergistic benefits derived from the combination of multiple modalities. 
Late fusion approaches may report limited performance in scenarios where the interactions between modalities significantly contribute to overall performance.

Indeed, training multiple branches to produce modality-specific decisions is highly advantageous in mitigating the modality collapse issue commonly encountered in early and intermediate fusion approaches. 
Specifically, late fusion encourages the preservation of unique information inherent to each modality \cite{holste2021end, yala2019deep} by making independent decisions based on modality-specific representations.
By producing modality-specific decisions, it promotes a more balanced fusion of modalities, avoiding the dominance of a single modality and enhancing the utilization of the complementary information provided by different modalities.
Thus, the modality collapse issue can be alleviated by ensuring that the valuable knowledge in each modality is appropriately captured and integrated during the fusion process. 
Overall, multiple branches in late fusion can capture the distinct characteristics of each modality more effectively, leveraging the unique information present in each modality and building a nuanced understanding of multimodal data.

\subsubsection{Multi-level Fusion}
Different fusion strategies of multimodal data in precision oncology research offers several benefits and limitations. 
Researchers have been exploring early, intermediate, and late fusion strategies to leverage the advantages of each while minimizing their drawbacks.
For example, Zhuang et al. \cite{zhuang20223d} conducted a study using multi-sequence MRI images, dividing them into distinct T1-T1ce and T2-FLAIR groups. 
They concatenated the multimodal data within each group at an early level and used separate encoders to extract multimodal representations for each group. 
These representations were integrated using a cross-modal interaction module, known as intermediate fusion.
This early-intermediate fusion strategy is particularly suitable for the coexistence of heterogeneous and homogeneous data modalities.
In addition to early-intermediate fusion, a combination of intermediate and late fusion strategies also gained attention in recent studies \cite{he2020feasibility, li2022adaptive}, enabling a sophisticated integration of multimodal decisions or underlying features simultaneously. 
It involves modeling intricate multimodal interconnections at the intermediate level, resulting in a multimodal decision, which can then be aggregated with modality-specific decisions to generate the final decision.
Notably, intricate multimodal interconnections can be modeled at the intermediate level, producing multimodal decisions, which can be aggregated with modality-specific decisions to produce the final decision.
Furthermore, the modality collapse issue can be effectively addressed by incorporating modality-specific decision modules, ensuring that the unique information from each modality is appropriately captured and preventing the overshadowing of any modality.
The multi-level fusion strategy enables researchers to effectively capitalize on the strengths of different fusion strategies and facilitate a deeper understanding of multimodal data and informed decision-making.

\subsubsection{Comparative Analysis.}
In the field of multimodal data integration for precision oncology, four fusion strategies have emerged: early, intermediate, late, and multi-level fusion. 
Specifically, early fusion captures the interactions between modalities from the beginning, requiring a simpler architecture to integrate multimodal data. 
However, it faces challenges in terms of plain multimodal interconnections and the modality collapse issue. 
Moreover, intermediate fusion processes modalities separately and allows more flexible integration of multimodal data.
However, it comes with increased computational complexity and risks the occurrence of modality collapse issues.
Furthermore, late fusion employs modality-specific branches to process each modality independently, incorporate modality-specific decisions, and mitigate the modality collapse issue. 
Nonetheless, it introduces higher architectural complexity, increased computational demands, and difficulties in modeling intricate multimodal interconnections. 
At last, the exploration of multi-level fusion, which integrates multimodal knowledge at different levels, holds promise for leveraging the advantages of various fusion strategies. 
Although many recent works \cite{liu2023m, zhuang20223d, zheng2023multi, zhang2017convolutional} have compared these fusion strategies, the underlying datasets are typically small-scale and less representative, leading to varying conclusions.
Therefore, there is currently a lack of a comprehensive and reliable comparison between these fusion strategies, particularly when considering multimodal data in the context of precision oncology.
In conclusion, the field of multimodal data integration in precision oncology offers different fusion strategies, each with its unique strengths and challenges, necessitating careful consideration of the specific conditions and requirements to achieve optimal results.

\begin{figure*}[!t]
\centering
\includegraphics[width=0.99 \textwidth]{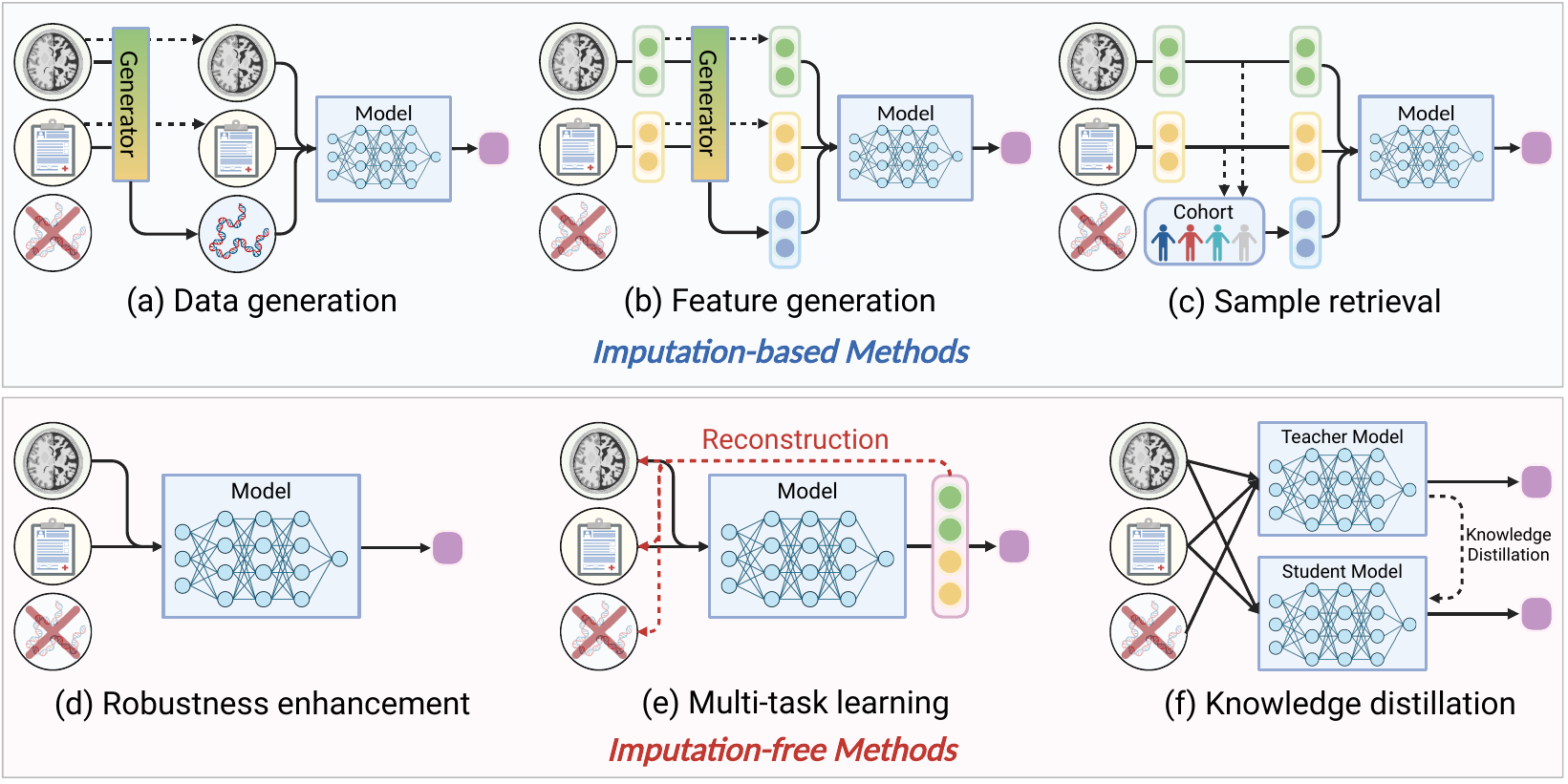}
\caption{Fusion strategies for incomplete date, including imputation-based methods and imputation-free methods.}
\label{fig:missing}
\end{figure*}

\subsection{Integration of Incomplete Data} Multimodal learning has emerged as a highly promising approach for acquiring comprehensive information about cancer patients by harnessing the power of diverse data sources. However, in the realm of clinical applications, the assumption of complete access to all modalities for fusion is often unattainable. It is a common occurrence to encounter missing data in one or more modalities, stemming from various factors including limitations in data collection, privacy concerns surrounding data sharing, and technical challenges associated with data acquisition. The presence of incomplete multimodal data poses a significant challenge to the performance of multimodal fusion models, and it can even lead to the failure of previously established fusion methods that heavily rely on complete modality data. Consequently, researchers have dedicated substantial efforts to address this critical issue within the domain of precision oncology. While previous studies and surveys such as~\cite{zhou2023literature, shah2023survey} have provided valuable insights into the integration of incomplete multimodal data, their focus has been primarily confined to methods applicable to homogeneous multimodal data, such as multi-sequence MRI data. These studies fail to offer a comprehensive overview of the existing techniques that can be applied to both homogeneous and heterogeneous multimodal data. Hence, in this section, we aim to bridge this gap by providing a comprehensive review of the existing methods employed for integrating incomplete multimodal data within the context of precision oncology, as shown in Fig. \ref{fig:missing}. Our investigation will shed light on the strengths, limitations, and potential applications of these techniques, thereby contributing to the advancement of multimodal fusion research in precision oncology.

\subsubsection{Imputation-based Methods} Intuitively, the most straightfoward solution to address the issue of missing modalities is to employ imputation techniques, which involve filling in the missing modalities using information from the observed modalities. Imputation-based methods can be further categorized into three subcategories: \textbf{i) imputation via data generation}, \textbf{ii) imputation via feature generation}, and \textbf{iii) imputation via sample retrieval}.

In the case of imputation via data generation, researchers commonly employ generative models such as generative adversarial networks (GANs) and variational autoencoders (VAEs). These models serve the purpose of synthesizing the missing modalities by leveraging the information present in the observed modalities. Subsequently, the generated modalities are combined with the observed ones through multimodal fusion techniques, enabling downstream tasks to be performed. Typically, all available modalities are integrated to learn a shared modality-invariant latent representation, which effectively captures the underlying data distribution. This latent representation is then utilized as a generation condition for the generative model to synthesize the missing modalities~\cite{chartsias2017multimodal, sharma2019missing, zhou2020hi, wu2023collaborative, chen2024modality}. Essentially, the missing information of incomplete multimodal data is modality-specific components that are not shared across all modalities. Consequently, the objective of generative model is to capture the modality-specific information embedded within the missing modalities. To enhance the quality of the generated modalities, certain studies explicitly incorporate a feature disentanglement scheme, which decomposes the available modalities into modality-invariant and modality-specific components~\cite{shen2020multi, peng2021multi}, thereby exploring and completing the possible modality-specific information contained within the missing modalities. To handle various possible missing modality cases, these early studies often need to train multiple generative networks. To improve the efficiveness of generation process and reduce the required computational resources, some studies have proposed to train a unified generative network that can handle multiple missing modality cases~\cite{hamghalam2021modality, dalmaz2022resvit, yang2023learning, yuan2023rethinking}. Notably, the diffusion model, renowned for its success in image generation and inpainting tasks, has also motivated the development of diffusion-based imputation methods for multimodal data\cite{meng2024multi}. Nonetheless, it is worth highlighting that the majority of methods in this category primarily focus on multi-sequence MRI data, which represents a typical example of homogeneous multimodal data. When confronted with highly heterogeneous data, the generative model may encounter challenges in effectively learning the complex data distribution, thereby resulting in suboptimal imputation performance.

To utilize generative models for synthesize missing modalities in highly heterogeneous multimodal data, some researchers have proposed imputation via feature generation methods~\cite{huang2021aw3m, hou2023hybrid, wang2023multi, ting2023multimodal, jiao2023gmrlnet}. These methods focus on extracting features from observed modalities and leveraging them to impute the missing modalities through the training of feature-level generative models. Unlike the aforementioned imputation techniques that operate at the data level, imputation via feature generation methods specifically target the generation of low-dimensional feature representations for the missing modalities. By extracting relevant features from the available modalities, these methods aim to estimate the latent representation of the missing modalities. This feature-level generation approach offers several advantages in effectively handling the complexities posed by highly heterogeneous multimodal data. One key advantage is the elimination of the need for the generative model to capture the high-dimensional data distribution, which can be computationally demanding and challenging in such heterogeneous settings. Instead, these methods prioritize the capture and estimation of low-dimensional feature representations, which often prove to be more feasible and effective in addressing the missing modality problem. It is important to highlight that imputation via feature generation methods is applicable to both homogeneous and heterogeneous multimodal data, presenting a versatile solution for various scenarios. However, a notable challenge arises from the abstract nature of the generated features. Consequently, the direct interpretation of these features becomes a non-trivial task. Moreover, evaluating and controlling the quality of the generated features pose additional difficulties in this context.

The aforementioned methods generally rely on the generative model to synthesize the missing modalities, which may be computationally expensive and sensitive to the scale of training set, suffering from the issue of mode collapse. These methods aim to address missing modalities by aggregating compensation information from similar samples in the training set~\cite{chen2020hgmf, zhang2022m3care}. Specifically, given the incomplete data, the sample retrieval methods leverage observed information to calculate the similarity among the inference sample and the samplesin the training set. Subsequently, the retrieved samples, possessing the desired modalities, are then used to fill in the missing modalities. To enhance the efficiveness and effectiveness of the retrieval process, some studies have proposed the utilization of learnable prototypes or prompts as representations for the samples in the training set~\cite{chen2023towards, wang2024mgiml}. Then, attention mechanisms are employed to aggregate the compensation information from the prototypes, which are learned from the entire training set. Nevertheless, the performance of sample retrieval methods heavily depends on the quality of the retrieved samples, which may not always accurately represent the true underlying data distribution and can introduce bias into the fusion model. Achieving precise retrieval of similar samples often necessitates predefined metric learning or similarity measurement, which can be impractical in certain scenarios. Furthermore, due to the limited number of samples in the training set, the modalities imputated by sample retrieval methods may lack the necessary diversity to fully capture the underlying data distribution. These limitations pose challenges that need to be addressed in order to enhance the effectiveness and reliability of sample retrieval methods.

Indeed, imputation-based methods offer valuable approaches to address the challenges associated with missing modalities in multimodal data. The choice of method depends on factors such as the characteristics of the data and the specific requirements of the application. Imputation via data generation methods is well-suited for homogeneous multimodal data, as it excels in synthesizing missing modalities. Imputation via feature generation methods, on the other hand, provides a versatile solution that can be applied to both homogeneous and heterogeneous multimodal data. This approach focuses on estimating low-dimensional feature representations for the missing modalities. Lastly, imputation via sample retrieval methods offers an efficient alternative by aggregating compensation information from similar samples in the training set. Despite their strengths, these methods also face several challenges, such as the scalability of the generative model, the interpretability of the generated features, and the diversity of the retrieved samples. Future research should aim to enhance the performance, interpretability, and scalability of imputation-based methods in addressing missing modalities, particularly in highly heterogeneous multimodal data settings.

\subsubsection{Imputation-free Methods} In contrast to imputation-based methods, imputation-free methods provide a more efficient and flexible solution for handling missing modalities in multimodal data. These methods directly leverage the observed modalities to perform multimodal fusion without imputing the missing modalities using sample information. Imputation-free methods can be further categorized into three subcategories: \textbf{i) robustness enhancement}, \textbf{ii) multi-task learning}, and \textbf{iii) knowledge distillation}.

Robustness enhancement methods aim to improve the robustness of multimodal fusion models to missing modalities, allowing the models to adapt to their presence. These methods generally utilize sophisticated fusion strategies and modules that are designed to be less sensitive to missing modalities, effectively enhancing the robustness of multimodal fusion models~\cite{havaei2016hemis, ning2021relation, ding2021rfnet, liu2022moddrop, wu2023multimodal, liu2023sfusion, 10288381}. Some of these models focus on capturing modality-invariant features shared across modalities, while others aim to directly integrate all available information from the observed modalities. In the domain of deep learning, specialized frameworks such as transformers and graph neural networks exhibit insensitivity to the dimensionality of input data, making them successful tools for fusing incomplete multimodal data~\cite{zhao2022modality, zhang2022mmformer, shi2023m, qiu2023modal, zhang2024tmformer}. Undoubtedly, robustness enhancement methods offer flexibility and efficiency in handling diverse cases of missing modalities. However, it is important to note that these methods may overlook the modality-specific information present in the available modalities, which can limit their performance in scenarios where modality-specific information is crucial for downstream tasks.

To ensure that robustness enhancement models capture the modality-specific information of each observed modality, researchers have introduced multi-task learning strategies that employ auxiliary tasks to encourage the model to learn complementary information from the observed modalities. One widely used auxiliary task is the reconstruction task, which aims to reconstruct the input data or features from the fused representation, effectively enforcing the model to retain essential information from all observed modalities~\cite{van2018learning, dorent2019hetero, zhou2020brain, zhou2021latent, cui2022survival, liu2023learning, zhou2023feature, liu2023m3ae}. Compared to robustness enhancement methods, multi-task learning methods implicitly capture modality-specific information from the observed modalities, further improving the performance of robustness enhancement models in handling missing data. However, these methods may still struggle to model the modality-specific information contained in the missing modalities.

To effectively capture the modality-specific information of missing modalities, knowledge distillation methods offer a promising solution. These methods transfer knowledge from a teacher model trained on complete multimodal data to a student model trained on incomplete data. Knowledge distillation can be performed at different levels, including feature-level distillation, relation-level distillation, and response-level distillation. The distinction lies in the granularity of the distilled knowledge, with feature-level distillation focusing on low-level features, relation-level distillation capturing high-level relations between features, and response-level distillation targeting the final prediction responses. Existing studies either adopt one of these distillation strategies~\cite{ ning2022mutual, konwer2023enhancing, wang2023learnable, qiu2023scratch} or combine multiple distillation strategies~\cite{hu2020knowledge, wang2021acn, vadacchino2021had, yang2022d, azad2022smu, karimijafarbigloo2024mmcformer}. By leveraging the distilled knowledge, the student model can achieve comparable performance to the teacher model, even in the presence of missing modalities.

Imputation-free methods offer a more efficient and flexible solution for handling missing modalities in multimodal data. These methods directly leverage the observed modalities to perform multimodal fusion without imputing the missing modalities. Robustness enhancement methods focus on improving the robustness of multimodal fusion models to missing modalities, while multi-task learning strategies encourage the model to learn complementary information from the observed modalities. Knowledge distillation methods transfer knowledge from a teacher model trained on complete multimodal data to a student model trained on incomplete data. However, they lack the ability to explicitly model the modality-specific information contained in the missing modalities. This limitation may impact their performance in certain scenarios.

\subsubsection{Comparative Analysis} In summary, both imputation-based and imputation-free methods offer valuable approaches to address the issue of missing modalities in multimodal fusion. Imputation-based methods focus on compensate the missing modalities using information from the observed modalities, while imputation-free methods directly leverage the observed modalities to perform multimodal fusion without imputing the missing modalities. The former can complete the missing modalities to some extent, but may suffer from the issue of mode collapse and the difficulty in capturing the complex data distribution. The latter provides a more efficient and flexible solution for handling missing modalities, but may struggle to generalize well to incomplete data if the shared representation or distilled knowledge fails to capture the essential information of missing modalities. It is important to note that there is no one-size-fits-all solution for addressing the challenges posed by missing modalities in multimodal fusion. Future research should aim to comprehensively analyze the strengths and limitations of both imputation-based and imputation-free methods. Additionally, exploring novel techniques that combine the advantages of these approaches can help enhance the robustness and generalization capabilities of multimodal fusion models when dealing with incomplete data. By considering the strengths and weaknesses of each method and developing novel techniques that leverage their respective advantages, researchers can advance the field of multimodal fusion and effectively address the challenges associated with missing modalities in diverse real-world applications.

\section{Applications of Multimodal Data Integration}
\label{sec:application}
Multimodal data integration has facilitated various applications of precision oncology, including early assessment, diagnosis, prognosis, and biomarker discovery.

\subsection{Early Assessment}
\subsubsection{Risk Stratification}
The goal of risk stratification is to evaluate an individual's risk of developing cancer in the early future based on various factors, including personal medical history, lifestyle choices, genetic predisposition, \textit{etc}. 
Traditional risk stratification tools \cite{RATool} simply estimate the risk using personalized information, such as race, age, diet, and medical history.
Recent research \cite{yala2019deep} found that risk stratification for breast cancer leveraging multi-view mammography images is significantly more accurate than traditional Tyrer-Cuzick (version 8) model \cite{tyrer2004breast}, while combining them obtains the optimal accuracy.
It indicates that multimodal data integration is a promising tool for breast cancer risk stratification and may be feasible for other cancers.

\subsubsection{Screening}
Cancer screening helps detect cancer early, improving the chances of successful treatment by identifying abnormal tissue before symptoms manifest and allowing for more effective intervention.
It relies on specific examinations to detect cancer in individuals, such as physical exams, laboratory tests, or imaging procedures.
Various data modalities produced in these examinations imply a great potential for multimodal data integration to improve screening accuracy.
For instance, Wu et al. \cite{8861376} showcased a remarkable method that seamlessly consolidates multiple mammography images, achieving a good accuracy on par with experienced radiologists.
Several studies have drawn similar conclusions regarding the screening of patients with diverse cancer types, such as prostate \cite{rossi2020multi}, ovarian \cite{xiang2024development}, breast \cite{liao2019automatic}, and upper gastrointestinal (UGI) \cite{ding2020scnet} cancer.
In summary, these studies highlight that integrating multimodal information in cancer screening produces more valuable insights than relying solely on a single modality.

\subsubsection{Detection}
Lesion detection aims at identifying the presence and location of lesions within the human body. 
It typically involves the use of various imaging techniques \cite{alyafeai2020fully}, such as MRI, CT, or PET, to identify abnormal growths or masses that may indicate the presence of a tumor.
For example, Kumar et al. \cite{kumar2019co} proposed a co-learning method to detect and segment tumors of lung cancer simultaneously.
To reduce the annotation burdens, researchers \cite{wang2018automated, yang2017co} utilized weakly-supervised methods to localize prostate tumors using sample-level labels.
They demonstrated the feasibility of automatically and accurately locating lesions with a small amount of annotation effort.

\subsection{Diagnosis}
\subsubsection{Segmentation}
Lesion segmentation refers to the process of outlining the boundaries of a tumor on medical imaging scans, \textit{e.g.}, CT, MRI, or PET images.
It empowers clinicians with more valuable information about the tumor's location, size, shape, and relationship with surrounding structures and organs, facilitating personalized treatment planning and decision-making.
Segmentation has garnered significant attention in recent years, largely attributed to the availability of high-quality datasets.
For example, the brain tumor segmentation (BraTS) challenge series \cite{menze2014multimodal} established a good benchmark consisting of multi-sequence MRI images and pixel-level annotations of the enhancing tumor (ET), the tumor core (TC), and the whole tumor (WT).
Extensive efforts \cite{yue2023adaptive, lin2023ckd, ma2018concatenated, ding2021mvfusfra, zhu2023brain, nie20183} have been dedicated to improving segmentation accuracy on this benchmark through the development of effective multimodal data integration techniques.
Besides, some works focus on segmenting lung \cite{zhou2023coco, xiang2022modality, podobnik2023multimodal}, head \& neck \cite{shi2023h}, prostate \cite{zhang2021cross}, colon \cite{lin2023lesion}, liver \cite{mo2020multimodal} tumors in PET-CT image pairs or multi-sequence MRI, as well as brain lesions \cite{zhang2024robust, chen2018voxresnet, zhuang2021aprnet} on other datasets \cite{mendrik2015mrbrains}.
In conclusion, multimodal data integration has a significant impact on improving segmentation accuracy, as evidenced by numerous successful efforts.

\subsubsection{Subtyping}
Cancer subtyping refers to the process of categorizing tumors into distinct subgroups based on their molecular, genetic, histological, or clinical characteristics. 
These subtypes represent different manifestations of cancer, each with unique biological features, behavior, and response to treatments. 
Subtyping holds paramount significance in the realms of treatment selection, outcome prediction, patient care guidance, and patient support.
Recently, several works \cite{kim2023heterogeneous, fang2024dynamic, alwazzan2024foaa, wang2021mogonet} leveraged advanced multimodal data integration techniques to build deep models for predicting breast cancer subtypes using multi-omics data.
Zheng et al. \cite{zheng2023multi} cooperated with both feature- and label-level confidence learning for cancer subtyping.
Moreover, Han et al. \cite{han2022multimodal} integrated multimodal data based on the estimated modality-specific informativeness scores.
The achievements of the aforementioned works underscore the importance of multimodal data integration for cancer subtyping.

\subsubsection{Grading}
Tumor grading is a process that assesses the cellular characteristics of cancer cells and determines the degree of abnormality or aggressiveness of a tumor. 
It involves examining tumor tissue samples under a microscope and assigning a grade based on specific histological features. 
It is essential for treatment decision-making, prognosis estimation, cancer monitoring, and effective communication with patients. 
Dozens of grading methods have been proposed for breast \cite{li2023msa}, glioma \cite{9478224}, prostate \cite{lara2020multimodal}, bladder \cite{8911262}, and hepatocellular carcinoma \cite{li2022adaptive} cancers.
For example, Zhang et al. \cite{8911262} predicted the grades of bladder cancer patients using pathology images and corresponding reports from clinicians.
They found that text information can improve grading performance by providing valuable clinical knowledge.
Cancer grading models, harnessing the power of multimodal data integration, hold profound implications for the advancement of precision oncology.

\subsubsection{Metastasis Prediction and Detection}
Metastasis refers to the identification of cancers that spread from their primary sites to distant organs or tissues in the body, impacting cancer staging and the choice of therapeutic interventions \cite{hou2023deep, hou2021integration}.
Metastasis prediction \cite{zheng2020deep} aims to estimate the likelihood of metastasis occurring in cancer patients, while metastasis detection methods \cite{10155265} are designed to identify the presence of metastasis in the given input data.
Hu et al. \cite{hu2023multi} leveraged the graph models to explore the relations between different features to detect lymph node metastasis (LNM). 
Qiao et al. \cite{qiao2022breast} effectively integrated MRI and US images by an explicit knowledge decomposition to jointly predict LNM, histological grade, and Ki-67 protein expression levels.
They also found that the modality-shared features precisely focus on tumor regions, extracting more tumor-related characteristics and improving the model's interpretability.
It suggests that multimodal data integration provides more precise information for metastasis prediction and detection, facilitating cancer staging and treatment planning. 

\subsubsection{Staging}
Staging refers to the process of determining the extent and spread of cancer within a patient's body. 
Existing staging systems, \textit{e.g.}, TNM system \cite{denoix1946enquete}, combine the extent of the tumor (T), the extent of spread to the lymph nodes (N), and the presence of metastasis (M) to classify cancers. 
The specific staging criteria may vary for different types of cancer, and clinicians rely on established guidelines and staging systems specific to each cancer type for accurate staging.
Multiple multimodal data integration-based staging models have been proposed recently. 
For example, Toney et al. \cite{toney2014neural} attempted to predict the nodal stage for non-small cell lung cancer by integrating CT and PET images.
Recently, Zhou et al. \cite{zhou2023rfia} leveraged endoscopic and pathology images to classify the stage of oesophageal cancers.
Multimodal data integration has shown promising potential for improving staging performance for a variety of cancers.

\subsection{Prognosis}
\subsubsection{Treatment Response Prediction}
Treatment response prediction refers to the estimation on how a patient will respond to a specific treatment or intervention. 
It involves using various factors, such as tumor characteristics, biomarkers, and imaging data, to assess the likelihood of a favorable response to a particular treatment. 
Treatment response prediction is valuable for clinicians as it helps guide treatment decision-making, optimize therapy selection, and improve patient outcomes.
For example, integrating CT, pathology, and genomics \cite{vanguri2022multimodal, boehm2022multimodal} for predicting treatment response in non-small cell lung and ovarian cancer patients can provide a quantitative rationale for clinicians.
More works \cite{anagnostou2020multimodal, jin2021predicting} for treatment response prediction indicate the significance of multimodal data integration.

\subsubsection{Survival Analysis}
Survival analysis is a statistical method used to analyze data related to the time until the occurrence of an event, \textit{i.e.}, death in survival analysis.
It provides valuable insights into the prognosis and outcomes of patients, allowing clinicians to make informed decisions about treatment options, monitor disease progression, and evaluate the effectiveness of therapeutic interventions.
The Cancer Genome Atlas (TCGA) program \cite{tcga} provides a wealth of clinical, imaging, and omics data, as well as follow-up records for patients with different cancers, significantly contributing to the development of multimodal data integration for survival analyses.
In these years, we have witnessed a large number of multimodal models \cite{zheng2022multi, li2022hfbsurv, jaume2023modeling, chen2020pathomic, nie20163d, qayyum20223d, tan2022multi, wu2023camr, zhang2022deep} for survival analysis using pathology and omics data.
These models hold promise in advancing our understanding of patient prognosis, guiding personalized treatment decisions, and ultimately improving treatment outcomes for individuals with cancer.

\subsubsection{Recurrence Prediction}
Recurrence prediction involves estimating the probability of cancer returning in patients who have received cancer treatment.
It involves analyzing various factors, such as tumor characteristics and treatment response, to assess the probability of the cancer returning after an initial remission. 
Recurrence prediction is valuable for clinicians as it helps guide surveillance strategies, inform treatment decisions, and optimize long-term management of patients \cite{nguyen2022attentive}.
Tang et al. \cite{tang2024new} identified high-risk recurrence after hepatic resection of colorectal cancer liver metastases using multi-sequence MRI images.
Gui et al. \cite{gui2023multimodal} developed a novel model integrating clinical, genomic, and histopathological data to improve the predictive accuracy for localized renal cell carcinoma recurrence.
These studies highlight the effectiveness of integrating multimodal data for predicting cancer recurrence.

\subsubsection{Tumor Growth Prediction}
Quantitatively characterizing the tumor's spatial-temporal progression is valuable in staging tumors and designing optimal treatment strategies.
Tumor growth not only relies on the properties of cancer cells but also depends on dynamic interactions between cancer cells and their constantly changing microenvironment. 
The complexity of the cancer system motivates the study of tumor growth using multimodal data.
For instance, Liu et al. \cite{liu2014patient} and Zhang et al. \cite{zhang2017convolutional} presented patient-specific tumor growth prediction models based on longitudinal dual-phase CT and PET imaging data.
These studies have provided valuable insights for tumor staging and treatment planning by analyzing tumor growth patterns and cell interactions using multimodal data.

\subsection{Biomarker Discovery}
Biomarkers are measurable indicators that can be used to detect the presence of cancer, predict its prognosis, monitor its progression, or evaluate the response to treatment.
There are many works \cite{nabbi2023multimodal,carneiro2015weakly, wei2023multi, shi2023novel, braman2021deep} attempted to analyze biomarker related to cancer diagnosis and prognosis based on multimodal data integration.
For imaging data, the Grad-CAM technique \cite{selvaraju2017grad} provides a powerful tool for analyzing which regions in images have high responses to the model's decisions \cite{chen2021multimodal, chen2020pathomic, chen2022pan, zhou2023texture}.
Moreover, genome-wide association studies (GWASs) \cite{claussnitzer2020brief, tam2019benefits, wang2005genome} aim to identify genomic variants that are statistically associated with cancer susceptibility \cite{amos2008genome, farashi2019post, wu2018transcriptome}.
Furthermore, Shapley's additive interpretation (SHAP) \cite{lundberg2017unified} is widely used in clinical records to understand the contribution of each input feature to model predictions.
In addition, the elucidation of cross-modal attention has emerged as a valuable technique for deciphering the intricate interconnections between different modalities, enabling a deeper understanding of how information from diverse modalities interacts with each other.
Rather than obsessing over the opacity of AI models, some researchers argue that it is crucial to emphasize the importance of rigorous validation through randomized clinical trials \cite{ghassemi2021false}.
Prospective trials enable us to thoroughly assess AI models under real-world conditions, compare their performance to standard-of-care practices, evaluate how clinicians interact with the AI tool, and determine the most effective way to integrate the models into the clinical workflow without causing disruption \cite{lipkova2022artificial}.

\section{Challenges and Future Directions}
\label{sec:challenge}
\subsection{Robust Learning with Imperfect Data}
Training multimodal AI models with strong robustness exhibits a strong need for large-scale and high-quality datasets.
However, collecting such datasets poses a significant challenge in the medical field, particularly in precision oncology, where the required examinations for cancer diagnosis and prognosis vary depending on the personalized health conditions of patients.
In addition, collecting cancer diagnostic and prognostic labels from patients is very laborious, for example, survival status usually takes months or even years to collect.
These issues suggest that the collected data is prone to imperfections, such as missing modalities or labels, which can significantly influence the effectiveness and generalizability of AI models.

In cases where samples have missing modalities, imputation-based methods \cite{chartsias2017multimodal, sharma2019missing, zhou2020hi, wu2023collaborative, chen2024modality} can be employed to complete the missing modalities by either generative models or retrieval-based methods.
Generative models like Generative Adversarial Networks (GANs) and diffusion models have garnered significant attention and have proven successful in various precision oncology applications.
Meanwhile, retrieval-based methods also hold the potential to address missing modalities by leveraging similarities between samples to retrieve relevant modalities.
Another viable option is imputation-free methods \cite{havaei2016hemis, ning2021relation, ding2021rfnet, liu2022moddrop, wu2023multimodal, liu2023sfusion, 10288381}, which aim to maximize the utilization of available modalities to mitigate the potential performance degradation and enhance the robustness of AI models when dealing with missing modalities. 
Various techniques such as representation learning, multi-task learning, and knowledge distillation can be employed to enhance the model's robustness in the presence of incomplete modalities, without the need for imputation.
Both of the above approaches demonstrate promising potential for addressing missing modalities in clinical scenarios, but which one is better remains under-explored.

In cases where the labels are unavailable for some samples, label-efficient learning techniques such as weakly- or semi-supervised learning mitigate the dependence on fully labeled data.
These techniques empower AI models to leverage samples with weak or even no labels, thereby enhancing their performance and capabilities.
By leveraging these approaches, models can effectively utilize more samples, enabling the learning process to benefit from a broader range of available information and reduce the need for extensive labeling efforts.
Furthermore, extensive unavailable labels may result in a small-scale training dataset.
In this case, few-shot learning is a valuable approach that can address the challenge of limited labeled data. 
In few-shot learning, models are trained to recognize patterns and extract relevant features from a limited number of labeled samples, enabling efficient adaptation and generalization to unseen samples.
In addition, federated learning can be adapted to train models collaboratively across multiple decentralized devices without the need to share raw data. 
This approach has the advantage of increasing the scale of the training set while maintaining data privacy, thereby enhancing the robustness of the models.
Overall, the utilization of the above techniques depends on the specific requirements and constraints of the clinical scenario.

\subsection{Effective Integration of Heterogeneous Modalities}
Modality heterogeneity, the variations in information presented across different modalities, can manifest in terms of data formats, scales, resolutions, or even semantics, posing challenges for effective multimodal integration and analysis.
The presence of heterogeneity not only complicates the fusion process but also introduces the risk of information loss or mismatch between modalities.
To address this issue, researchers have explored various techniques and methodologies, such as cross-modality representation learning \cite{xiang2022modality, yue2023adaptive, liu2023m} to enhance modality representations based on multimodal interconnections, semantic alignment methods \cite{zhou2023cross, wu2023camr, lara2020multimodal} to mitigate the semantic gaps between modalities, or knowledge decomposition \cite{10155265, qiao2022breast} to explicitly model distinct knowledge components for a comprehensive integration.
Moreover, the availability of knowledge quantification tools \cite{liang2023quantifying} is essential for accurately quantifying the nature (knowledge type) and extent (knowledge amount) of interactions between modalities, providing valuable insights into the underlying patterns, correlations, and dynamics within the multimodal data. 
These tools can enable us to design a more effective fusion strategy that can leverage the full potential of multimodal data and provide a solid foundation for evaluating and comparing different multimodal data integration models.
Furthermore, the foundation model for multimodal data integration can provide general and discriminative representations by learning from large-scale datasets, which is also a promising direction.
Overall, the above directions can unlock the potential of multimodal data, informing the design of a better fusion strategy, enabling more accurate analysis and decision-making, and driving progress in precision oncology.

Information redundancy presents challenges for AI models to effectively discern between task-relevant and irrelevant information.
When multiple modalities provide a huge amount of information, it becomes difficult for models to discern which pieces of information are truly informative for the given task.
Various techniques have been adopted to reduce information redundancy in multimodal data integration, such as cross-modal feature selection \cite{xu2023multimodal}, task-oriented dimensionality reduction \cite{xu2024label}, metric learning \cite{shao2023fam3l}, and information bottleneck \cite{zhang2024prototypical, fang2024dynamic} methods.
These techniques have been proven effective in eliminating redundant information, leading to improved performance and efficiency in multimodal data integration.
The emergence of information theory-based approaches \cite{zhang2024prototypical} presents a promising direction by providing a solid theoretical foundation and sophisticated tools for effectively handling redundancy, allowing researchers to advance the state-of-the-art performance.
In summary, further investigations on information theory hold the potential to advance our comprehension of cancer and significantly improve cancer diagnosis, prognosis, and treatment decision-making processes.

In addition, the expertise of healthcare professionals contains a wide range of valuable knowledge about cancers, which can enhance the multimodal representations and improve the model's performance on clinical applications.
By tailoring the expertise knowledge to individual characteristics such as patient demographics or genetic profiles, we can enhance the applicability of the expertise knowledge in personalized cancer care and treatment.
To utilize expertise knowledge, researchers have developed expert-driven modules \cite{lin2023ckd, zhang2017tandemnet} that integrate clinical guidelines and best practices with multimodal data analysis.
Furthermore, collaboration between clinicians and data scientists allows for the identification of relevant clinical factors, contextual information, and domain-specific knowledge to enhance the representation extracted by AI models \cite{hu2023multi, li2023lvit, shao2023characterizing}.
The collaboration of researchers and clinicians can bridge the gap between data-driven models and clinical practice, leading to accurate diagnoses, tailored treatment plans, and improved patient outcomes.

\subsection{Explainable and Trustworthy AI Models}
Gaining trust is crucial for clinicians and patients to accept diagnoses and treatment recommendations provided by multimodal AI models.
To achieve this goal, multimodal AI models must demonstrate transparency in their decision-making and multimodal interconnection processes. 
It involves extracting meaningful insights \cite{schulte2021integration}, identifying the contributing factors, and providing transparent explanations for the decisions made by the model.
To facilitate decision interpretation, researchers have proposed various techniques and methodologies. 
These include visualization methods \cite{chen2020pathomic,zhang2017mdnet} that provide intuitive representations of the multimodal data and decision-making process, feature importance analysis \cite{li2022adaptive} to identify the most influential features to the decision, and rule extraction techniques \cite{yan2024combiner} that extract interpretable rules from the integrated data.
Moreover, by analyzing the learned cross-modal attention \cite{xu2023multimodal, chen2022pan}, we can reveal the relationships and dependencies between different modalities, providing valuable insights into the intricate interactions and complementary nature of multimodal data.
Besides gaining trust, improving the interpretability of AI models can also help clinical developments.
For example, it can highlight the pivotal role of modifiable risk factors, such as Mediterranean lifestyle and physical activity, to the susceptibility of cancer.
Moreover, model interpretability can guide clinicians to discover new biomarkers that provide valuable insights into the presence and characteristics of cancer, enabling personalized and targeted approaches to diagnosis and treatment. 
Furthermore, the utilization of cross-modal interconnections can offer remarkable benefits by enabling the discovery of non-invasive alternatives, thereby reducing the need for extensive examinations and minimizing patient discomfort and pain.
Overall, enhancing the interpretability of AI models brings numerous benefits to both clinicians and patients in precision oncology.

\subsection{Efficient Processing with Limited Resource}
Efficient processing with limited resources poses a significant challenge in the context of multimodal data integration for precision oncology. 
The integration of diverse data modalities typically requires sophisticated fusion strategies and computational frameworks that can effectively extract and fuse information from multiple sources. 
As the volume and complexity of multimodal data continue to grow, there is a need for handling multimodal data efficiently, particularly when resource constraints, such as limited computational power or storage capacity.
Future directions in this field involve developing resource-efficient modules \cite{shi2023h}, leveraging techniques such as dimensionality reduction, and parallel processing to optimize the processing and analysis of multimodal data. 
Additionally, advancements in hardware-friendly modules \cite{tang2023ac2as} can also contribute to more efficient processing of multimodal data. 
Moreover, the development of knowledge distillation \cite{hu2020knowledge} can help focus computational resources on the most informative features of multimodal data, ensuring efficient processing without compromising accuracy and precision in clinical applications \cite{nie20163d, pereira2016brain, zhang2017convolutional}. 
Overall, addressing the challenge of efficient processing with limited resources is crucial in accelerating the multimodal data integration models into practical and scalable solutions for precision oncology applications.

\subsection{Cross-center Adaption and Evaluation}
Cross-center adaption and evaluation present an important concern in the realm of multimodal data integration for precision oncology. 
It is crucial for multimodal data integration models to extract robust and general features and be applied across diverse patient populations and healthcare settings. 
However, differences in data acquisition protocols, imaging equipment, and clinical practices among centers pose challenges in harmonizing clinical data. 
Future directions in this area may focus on addressing these challenges by developing standardized protocols and guidelines for data acquisition, annotation, and representation to ensure compatibility and interoperability across centers. 
This includes the establishment of data-sharing collaborations that promote the exchange of multimodal data while maintaining patient privacy and data security. 
Additionally, the development of transfer learning \cite{wang2022continual} and domain adaptation \cite{chen2022contrastive, nagabandi2018learning} techniques can enable the adaptation of models trained on data from one center to another, bridging the gap between different centers and facilitating the integration of multimodal data. 
Furthermore, the establishment of robust evaluation benchmark datasets \cite{menze2014multimodal, zhou2024benchmarking, ma2024multimodality} that encompass diverse patient cohorts and center-specific characteristics is crucial for assessing the performance and generalizability of multimodal models across centers. 
Overall, addressing the challenge of cross-center adaption and evaluation is essential for advancing the field of multimodal data integration in precision oncology and ensuring the translation of research findings into clinical practice.

\section{Conclusion}
\label{sec:conclusion}
The booming development of multimodal data integration in cancer research has provided unprecedented discovery and advancement of precision oncology practice.
In this paper, we review about 300 papers on multimodal data integration for precision oncology over the past decade.
Specifically, integrating multimodal data at intermediate or late levels has gained substantial attention, while the emergence of the multi-level fusion strategy facilitates a more effective method to unveil intricate multimodal interconnections.
In tackling samples with missing modalities, both imputation-based and imputation-free methods have demonstrated their effectiveness in various clinical applications. 
However, determining which type of method is superior remains an unsettled question, as the conclusions of these studies are heavily influenced by the specific data utilized.
Through the discussion of existing challenges, we provide valuable insights on future directions for advancing multimodal data integration in precision oncology.
As precision oncology continues to evolve, embracing the power of multimodal data integration will undoubtedly shape the future of cancer care, offering enormous potential for personalized medicine and transforming the lives of countless patients worldwide.

\bibliographystyle{IEEEtran}
\bibliography{egbib}

\clearpage


\def\BibTeX{{\rm B\kern-.05em{\sc i\kern-.025em b}\kern-.08em
    T\kern-.1667em\lower.7ex\hbox{E}\kern-.125emX}}
\markboth{\journalname, VOL. XX, NO. XX, XXXX 2017}
{Author \MakeLowercase{\textit{et al.}}: Preparation of Brief Papers for IEEE TRANSACTIONS and JOURNALS (February 2017)}

\makeatletter
\renewcommand \thesection{S\@arabic\c@section}
\renewcommand\thetable{S\@arabic\c@table}
\renewcommand \thefigure{S\@arabic\c@figure}
\makeatother

\setcounter{enumiv}{0}

\title{Supplementary Material for Multimodal Data Integration for Precision Oncology: Challenges and Future Directions}
\author{Huajun~Zhou, \IEEEmembership{Member,~IEEE,} Fengtao~Zhou, Chenyu Zhao, Yingxue~Xu, Luyang Luo, \IEEEmembership{Member,~IEEE,} Hao~Chen*, \IEEEmembership{Senior Member,~IEEE.}}

\maketitle

\setcounter{section}{0}
\section{Appendix}
\subsection{Review Inclusion Criteria}
This survey aims to investigate the current state of multimodal data integration techniques in the field of precision oncology and provide insights into the current challenges and potential future directions.
We performed a comprehensive analysis of the literature from 2014 to the present (2024 up to April) by searching on the Google Scholar database using the keywords ``multimodal'', ``multi-modal'', ``multi-modality'', ``multiple modalities'', ``missing modality'', or ``incomplete modality'' joint with ``cancer''. 
In the search results, we filter out the papers of low quality or have not been validated on precision oncology-related tasks.
Finally, our search identified around 300 manuscripts to support the analysis of multimodal data integration for advancing precision oncology.

\subsection{Multimodal Datasets for Precision Oncology}
We present a summary of multimodal datasets on precision oncology in Tab. \ref{tab:dataset}, offering rich resources for researchers and clinicians interested in exploring these valuable topics.
We also provide a brief introduction to these datasets in the following.

The Cancer Genome Atlas (TCGA) \citesupp{1cancer2012comprehensive} stands as a pioneering cancer genomics initiative that began in 2006, encompassing molecular characterization of more than 20,000 primary cancers and matched normal samples across 33 diverse cancer types. 
Over the next dozen years, TCGA collected over 2.5 petabytes of genomic, epigenomic, transcriptomic, and proteomic data. 
In addition to the comprehensive omics data, the TCGA database offers a wealth of supplementary information, including clinical data, pathology images, and more.
The data, which has already led to improvements in our ability to diagnose, treat, and prevent cancer, will remain publicly available for anyone in the research community to use.

The Multimodal Brain Tumor Segmentation (BraTS) \citesupp{1menze2014multimodal, 1deverdier20242024} challenge was launched at the 2012 MICCAI conference.
The latest BraTS 2024 challenge contains multimodal data from over 4,500 cases, including multi-institutional pre-operative baseline multi-parametric magnetic resonance imaging (MRI) scans, and focuses on the evaluation of state-of-the-art methods for the segmentation of intrinsically heterogeneous brain glioblastoma sub-regions in MRI scans. 
Furthermore, it presents newly proposed clinically relevant challenges, in a synergistic attempt to maximize the potential clinical impact of the innovative algorithmic contributions made by researchers. 
The scope extends further to address additional i) underserved populations (\textit{i.e.}, sub-Saharan Africa patients), ii) timepoints (\textit{i.e.}, pre- \& post-treatment), iii) tumor types (\textit{e.g.}, meningioma), iv) modalities (\textit{i.e.}, histology samples), v) clinical concerns (\textit{e.g.}, missing data), and iv) technical considerations (\textit{e.g.}, generalizability). 
In conclusion, the BraTS 2024 datasets describe a further contribution to the community of additional well-curated manually-annotated cases, comprising MRI scans from 4,000 previously unseen patients and 280,000 histology samples.

The Head and Neck organ-at-risk CT \& MR segmentation (HaN-Seg) challenge \citesupp{1podobnik2023han} comprises CT and T1-weighted MR images of 56 patients, which were deformably registered with the SimpleElastix registration tool, and corresponding curated manual delineations of 30 organs-at-risk.

The HEad and neCK TumOR (HECKTOR) 2022 challenge \citesupp{1oreiller2022head} has a primary objective centered around segmentation and outcome prediction utilizing PET and CT imaging modalities. 
This challenge involves approximately 845 cases, where 524 and 489 cases as the training sets, and 359 and 339 cases as the testing sets for head and neck primary tumors and lymph nodes segmentation (task 1) and recurrence-free survival prediction (task 2), respectively.
By leveraging the combined power of clinical records and PET/CT imaging, this challenge aims to advance the field's understanding of tumor segmentation techniques and predictive modeling for patient outcomes.

The autoPET dataset \citesupp{1gatidis2022whole} provides an annotated dataset of oncologic PET/CT studies for the development and training of machine learning methods and to help address the limited availability of publicly available high-quality training data for PET/CT image analysis projects.
It contains 501 consecutive whole-body FDG-PET/CT cases of patients with malignant lymphoma, melanoma, and non-small cell lung cancer (NSCLC) as well as 513 cases without PET-positive malignant lesions examined between 2014 and 2018 at the University Hospital Tübingen.

The Prostate Imaging: Cancer AI (PICAI) challenge \citesupp{1saha2023artificial} contains over 10,000 carefully curated prostate multi-sequence MRI exams to validate modern AI algorithms and estimate radiologists’ performance at CS PCa detection and diagnosis.
It primarily consists of two sub-studies: 1) an automatic evaluation on a multi-center, multi-vendor dataset and 2) international prostate radiologists perform a reader study using a subset of 400 scans from the hidden cohort.

Duke Breast Cancer MRI (DUKE) \citesupp{1saha2018machine} collected 922 biopsy-confirmed invasive breast cancer patients with breast cancer from a retrospective study of a decade years. 
Each case contains a nonfat-saturated T1-weighted sequence, a fat-saturated T1-weighted precontrast sequence, and mostly three to four post-contrast sequences. 
The dataset also provides non-imaging information such as demographics, treatments, tumor characteristics, recurrence, \textit{etc.}, which could help researchers implement multiple further tasks.

The CMMD database \citesupp{1cai2023online} provides two views of breast tumors for each patient: craniocaudal (CC) and mediolateral oblique (MLO) views, as well as four other clinical features that can support the imaging modality. 
Clinical features include the location of the lesion (as left or right breast), age, lesion subtype (such as mass, calcification, or both), and molecular subtypes like luminal A, luminal B, HER2 positive, and Triple-negative. 
The CMMD comprises 3,728 mammograms from 1,775 cases, where 481 of which are benign and 1294 malignant.

The Medical Segmentation Decathlon (MSD) \citesupp{1antonelli2022medical} is a biomedical image analysis challenge, in which algorithms compete in a multitude of both tasks and modalities to investigate the hypothesis that a method capable of performing well on multiple tasks will generalize well to a previously unseen task and potentially outperform a custom-designed solution.
It consists of over 2600 volumes of multi-sequence MRIs, including T1, post-contrast T1-weighted (T1Gd), T2, and Fluid Attenuated Inversion Recovery (FLAIR), and CT images for multiple cancer types, such as liver, brain, lung, prostate, \textit{etc}.

The MR Brain Segmentation (MRBrainS) dataset \citesupp{1mendrik2015mrbrains} consists of 7 sets of brain MR images (T1, T1 inversion recovery, and T2-FLAIR) with manual segmentation of ten brain structures. 
These manual segmentation have been made by experts in brain segmentation. 
The introduced MRBrainS18 challenge represents a notable advancement over its predecessor, the MRBrainS13 challenge, by incorporating pathology images into its evaluation framework. 
This addition allows for a more comprehensive assessment of brain imaging techniques and algorithms, enabling participants to address the challenges posed by both MRI and pathology data.

HAM10000 \citesupp{1tschandl2018ham10000} collected dermatoscopic images from different populations, acquired and stored by different modalities. 
The final dataset consists of 10015 dermatoscopic images which can serve as a training set for academic machine learning purposes. 
Cases include a representative collection of all important diagnostic categories in the realm of pigmented lesions: Actinic keratoses, intraepithelial carcinoma / Bowen's disease, basal cell carcinoma, benign keratosis-like lesions, dermatofibroma, melanoma, melanocytic nevi, and vascular lesions.

The SPIE-AAPM-NCI Prostate MR Classification (PROSTATEx) Challenge \citesupp{1armato2018prostatex} was held in conjunction with the 2017 SPIE Medical Imaging Symposium. 
It collects a retrospective set of prostate MR sequences, including T2-weighted (T2W), proton density-weighted (PD-W), dynamic contrast-enhanced (DCE), and diffusion-weighted (DW) imaging. 
It released a training set of cases in November 2016 that contained mpMRI scans of 330 prostate lesions from 204 patients along with spatial location coordinates, anatomic zone location, and known clinical significance of each lesion. 
Three weeks later the test set of cases was made available; the test set contained mpMRI scans of 208 prostate lesions from 140 patients with spatial location and anatomic zone, but the clinical significance information for these lesions was not included. 

Wisconsin Neurodevelopment Rhesus (WNR) dataset \citesupp{1young2017unc} contains longitudinal data from both structural and diffusion MRI images generated on a cohort of 34 typically developing monkeys from 2 weeks to 36 months of age. 
All images have been manually skull-stripped and are being made freely available via an online repository for use by the research community.

The Lung Image Database Consortium image collection (LIDC-IDRI) \citesupp{1armato2011lung} consists of diagnostic and lung cancer screening thoracic computed tomography (CT) scans with marked-up annotated lesions. 
It contains 1018 cases of images from a clinical thoracic CT scan and an associated XML file that records the results of a two-phase image annotation process performed by four experienced thoracic radiologists.
LUNA16 \citesupp{1murphy2009large} is a subset of the LIDC-IDRI dataset, containing 888 CT scans and corresponding annotations which were collected during a two-phase annotation process using four experienced radiologists. 
Combining these two datasets can construct a multimodal dataset for lung nodule classification.

The INbreast \citesupp{1MOREIRA2012236} dataset was created to provide a standardized and representative collection of mammograms for research and development purposes. 
It has a total of 115 cases (410 images) from which 90 cases are from women with both breasts affected (four images per case) and 25 cases are from mastectomy patients (two images per case), as well as corresponding manual annotations of the pectoral muscle boundary.
Moreover, clinical records for each patient are also provided, including BI-RADS scores, lesion annotations (masses, calcifications, asymmetries, and distortions), and medical reports.
This diversity allows researchers and practitioners to explore various aspects of breast cancer detection and diagnosis, such as computer-aided detection (CAD) systems, image analysis algorithms, and classification models.

The IXI dataset \citesupp{1IXIdataset} contains 566 scans from normal subjects and each scan has three modalities: T1, T2, PD-weighted, MRA, and Diffusion-weighted images (15 directions).

Digital Database of Screening Mammography (DDSM) \citesupp{1Michael2001digital} is a dataset with 2620 scanned film mammography cases. 
Each case contains two views, i.e., mediolateral oblique (MLO) view and craniocaudal (CC) view, for each breast, resulting in a total of 10,480 images. 
All cases are labeled as normal, benign, and malignant with pathological verification and manually generated ROI annotations (bounding boxes) for the abnormalities.

\begin{table*}[ht]
    \centering
    \setlength\tabcolsep{9pt}
    \caption{Summary of multimodal datasets for precision oncology. Datasets with * indicate variations in statistics across different years. Since we present the statistics for a single year only, for more detailed information, please refer to their respective websites.}
    \label{tab:dataset}
    \begin{tabular}{|c|c|c|c|c|c|}
    \hline
        \textbf{Dataset} & \textbf{Year} & \textbf{Modality} & \textbf{Cancer type} & \textbf{\# of Sample} & \textbf{Link} \\ \hline
        TCGA* \citesupp{1cancer2012comprehensive} & 2006-now & Pathology, Omics, Clinical records, MRI & Multiple & \textgreater 20000 & \href{https://www.cancer.gov/aboutnci/organization/ccg/research/structural-genomics/tcga}{Link} \\ \hline
        BraTS* \citesupp{1menze2014multimodal, 1deverdier20242024} & 2012-now & Multi-MRI & Brain & \textgreater 4500 & \href{https://zenodo.org/records/10978907}{Link} \\ \hline
        HaN-Seg \citesupp{1podobnik2023han} & 2023 & CT, MRI & Head \& Neck & 56 & \href{https://han-seg2023.grand-challenge.org/}{Link} \\ \hline
        HECKTOR* \citesupp{1oreiller2022head} & 2022 & CT, PET, Clinical records & Head \& Neck & 845 & \href{https://hecktor.grand-challenge.org/Data/}{Link} \\ \hline
        autoPET \citesupp{1gatidis2022whole} & 2022 & CT, PET, Clinical records & Multiple & 1014 & \href{https://www.cancerimagingarchive.net/collection/fdg-pet-ct-lesions/}{Link} \\ \hline
        PICAI \citesupp{1saha2023artificial} & 2022 & Multi-MRI, Clinical records & Prostate & $\sim$11000 & \href{https://pi-cai.grand-challenge.org/}{Link} \\ \hline
        DUKE \citesupp{1saha2018machine} & 2021 & MRI, Clinical records & Breast & 922 & \href{https://wiki.cancerimagingarchive.net/pages/viewpage.action?pageId=70226903}{Link} \\ \hline
        CMMD \citesupp{1cai2023online} & 2021 & MRI, Text & Breast & 1775 & \href{https://wiki.cancerimagingarchive.net/pages/viewpage.action?pageId=70230508}{Link} \\ \hline
        MSD \citesupp{1antonelli2022medical} & 2019 & Multi-MRI, Multi-CT & Multiple & $\sim$2600 & \href{https://registry.opendata.aws/msd/}{Link} \\ \hline
        MRBrainS* \citesupp{1mendrik2015mrbrains} & 2018 & Multi-MRI & Brain & 30 & \href{https://mrbrains18.isi.uu.nl/index.html}{Link} \\ \hline
        HAM10000 \citesupp{1tschandl2018ham10000} & 2018 & Image, Clinical records & Skin & 10015 & \href{https://dataverse.harvard.edu/dataset.xhtml?persistentId=doi:10.7910/DVN/DBW86T}{Link} \\ \hline
        PROSTATEx* \citesupp{1armato2018prostatex} & 2017 & Multi-MRI & Prostate & 344 & \href{https://prostatex.grand-challenge.org/}{Link} \\ \hline
        WNR \citesupp{1young2017unc} & 2017 & Multi-MRI & Brain & 34 & \href{https://www.nitrc.org/projects/uncuw\_macdevmri}{Link} \\ \hline
        LIDC-IDRI \citesupp{1armato2011lung} + LUNA16 \citesupp{1murphy2009large} & 2016 & CT, Clinical records & Lung & 888 & \href{https://luna16.grand-challenge.org/Data/}{Link} \\ \hline
        InBreast \citesupp{1MOREIRA2012236} & 2012 & Multi-mammogram & Breast & 115 & \href{https://www.kaggle.com/datasets/tommyngx/inbreast2012}{Link} \\ \hline
        IXI \citesupp{1IXIdataset} & - & Multi-MRI, Clinical records & Brain & 566 & \href{https://brain-development.org/ixi-dataset/}{Link} \\ \hline
        DDSM \citesupp{1Michael2001digital} & 2001 & Multi-mammogram & Breast & $\sim$2620 & \href{http://www.eng.usf.edu/cvprg/mammography/database.html}{Link} \\ \hline
    \end{tabular}
\end{table*}

\subsection{Full Lists of Reviewed Papers}
Our survey includes an extensive list of the reviewed papers discussed in our works, showcasing our commitment to thorough analysis. 
The integration of complete data (Tab. \ref{tab:fusion}) and incomplete data (Tab. \ref{tab:missing}) ensures transparency and provides valuable insights into the research landscape. 
Moreover, dozens of survey papers (Tab. \ref{tab:survey}) also provide valuable information for our comprehensive analysis.

\bibliographystylesupp{IEEEtran}
\bibliographysupp{egbib_sub}

\clearpage
\onecolumn
\begin{center}
\captionof{table}{Summary of reviewed papers on integrating complete multimodal data. SA: Survival analysis; SEG: Segmentation; DET: Detection; GP: genomic prediction; TGP: Tumor growth prediction; TRP: Treatment response prediction; RP: Recurrence prediction; MD: Metastasis detection; Path: Pathology; Omic: Omics Data; MMG: Mammography; EI: Endoscopy Imaging; DI: Dermoscopy Imaging; CR: Clinical Records; BCW: Breast Cancer Wisconsin.}
\tablefirsthead{\hline}
\tablehead{%
    \hline
    \multicolumn{2}{l}{\small\sl Continued from previous page}\\
    \hline}
\tabletail{%
    \hline
    \multicolumn{2}{l}{\small\sl Continued on next page}\\
    \hline}
\tablelasttail{\hline}

\begin{supertabular}{|l|l|l|l|l|l|l|}
\label{tab:fusion}
        \textbf{Approach} & \textbf{Cancer type} & \textbf{Application} & \textbf{Modality} & \textbf{Dataset} & \textbf{Publication} \\ \hline
        Toney et al. \citesupp{1toney2014neural} & Lung & Staging & PET, CT, CR & Private & Radiology 2014 \\ \hline
        Liu et al. \citesupp{1liu2014patient} & Pancreatic & TGP & CT, PET & Private & MIA 2014 \\ \hline
        Carneiro et al. \citesupp{1carneiro2015unregistered} & Breast & Screening & Multi-MMG & InBreast, DDSM & MICCAI 2015 \\ \hline
        Carneiro et al. \citesupp{1carneiro2015weakly} & H\&N & MCSU estimation & Multi-Path & humboldt & ICCV 2015 \\ \hline
        Nie et al. \citesupp{1nie20163d} & Brain & SA & Multi-MRI & Private & MICCAI 2016 \\ \hline
        Xu et al. \citesupp{1xu2016multimodal} & Cervical & Screening & EI, CR & Private & MICCAI 2016 \\ \hline
        Pereira et al. \citesupp{1pereira2016brain} & Brain & SEG & Multi-MRI & BraTS & TMI 2016 \\ \hline
        Li et al. \citesupp{1li2016brain} & Brain & SEG & Multi-MRI & BraTS & AI in Medicine 2016 \\ \hline
        Kamnitsas et al. \citesupp{1kamnitsas2017efficient} & Brain & SEG & Multi-MRI & BraTS & MIA 2017 \\ \hline
        Shen et al. \citesupp{1shen2017boundary} & Brain & SEG & Multi-MRI & BraTS & MICCAI 2017 \\ \hline
        Carneiro et al. \citesupp{1carneiro2017automated} & Breast & Screening & Multi-MMG & InBreast, DDSM & TMI 2017 \\ \hline
        Ge et al. \citesupp{1ge2017skin} & Skin & Subtyping & Multi-Image & Private & MICCAI 2017  \\ \hline
        Yao et al. \citesupp{1yao2017deep} & Multiple & SA & Path, Omic & TCGA & MICCAI 2017 \\ \hline
        Havaei et al. \citesupp{1havaei2017brain} & Brain & SEG & Multi-MRI & BraTS & MIA 2017 \\ \hline
        Yang et al. \citesupp{1yang2017co} & Prostate & DET & Multi-MRI & Private & MIA 2017 \\ \hline
        Zhang et al. \citesupp{1zhang2017tandemnet} & Bladder & Grading & Path, CR & BCIDR & MICCAI 2017 \\ \hline
        \multirow{2}{*}{Wang et al. \citesupp{1wang2018automated}} & \multirow{2}{*}{Prostate} & DET, Screening, & \multirow{2}{*}{Multi-MRI} & ProstateX, & \multirow{2}{*}{TMI 2018} \\ 
          &  & Registration &  & Private &  \\ \hline
        Ma et al. \citesupp{1ma2018concatenated} & Gliomas & SEG & Multi-MRI & BraTS & TMI 2018 \\ \hline
        Zhang et al. \citesupp{1zhang2017convolutional} & Pancreatic & TGP & CT, PET & Private & TMI 2018 \\ \hline
        Zhou et al. \citesupp{1zhou2018one} & Brain & SEG & Multi-MRI & BraTS & MICCAI 2018 \\ \hline
        Zhao et al. \citesupp{1zhao2018deep} & Brain & SEG & Multi-MRI & BraTS & MIA 2018 \\ \hline
        Pereira et al. \citesupp{1pereira2018adaptive} & Brain & SEG & Multi-MRI & BraTS & MICCAI 2018 \\ \hline
        Chen et al. \citesupp{1chen2018drinet} & Brain & SEG & Multi-MRI & BraTS & TIP 2018 \\ \hline
        Wang et al. \citesupp{1wang2018interactive} & Brain & SEG & Multi-MRI & BraTS & TMI 2018 \\ \hline
        Chen et al. \citesupp{1chen2019combination} & Prostate & Det & PET, CT, MRI, Path & Private & JNM 2019  \\ \hline
        Zhou et al. \citesupp{1zhou2019unet} & Brain & SEG & Multi-MRI & BraTS & TMI 2019 \\ \hline
        Akselrod et al. \citesupp{1akselrod2019predicting} & Breast & Screening & CR, MMG & Private & Radiology 2019   \\ \hline
        Chen et al. \citesupp{1chen2019dual} & Brain & SEG & Multi-MRI & BraTS & PR 2019 \\ \hline
        Chen et al. \citesupp{1chen20193d} & Brain & SEG & Multi-MRI & BraTS & MICCAI 2019 \\ \hline
        Yala et al. \citesupp{1yala2019deep} & Breast & Risk assessment & CR, MMG & Private & Radiology 2019 \\ \hline
        Kumar et al. \citesupp{1kumar2019co} & Lung & DET, SEG & PET, CT & Private & TMI 2019 \\ \hline
        Razzak et al. \citesupp{1razzak2018efficient} & Brain & SEG & Multi-MRI & BraTS & JBHI 2019 \\ \hline
        Cao et al. \citesupp{1cao2019joint} & Prostate & SEG & Multi-MRI & Private & TMI 2019 \\ \hline
        Zhang et al. \citesupp{1zhang2019multi} & Multiple & VQA & Image, Text & Private & ACMMM 2019 \\ \hline
        Vu et al. \citesupp{1vu2020question} & Breast & VQA & Path, Text & BreakHis & TMI 2020 \\ \hline
        Wang et al. \citesupp{1wang2020auto} & Breast & Screening & Multi-US & Private & MICCAI 2020 \\ \hline
        Chen et al. \citesupp{1chen2020combined} & Pancreatic & GP & Multi-MRI & Private & TMI 2020 \\ \hline
        Tang et al. \citesupp{1tang2020deep} & Brain & SA, Genotyping & Multi-MRI, CR & Private & TMI 2020 \\ \hline
        Wu et al. \citesupp{18861376} & Breast & Screening & Multi-MMG & Private & TMI 2020 \\ \hline
        Shao et al. \citesupp{1shao2019integrative} & Multiple & SA & Path, Omic & TCGA & TMI 2020 \\ \hline
        Gao et al. \citesupp{1gao2020mgnn} & Breast, Lung & SA & Multi-Omic, CR & cBioPortal & SIGIR 2020 \\ \hline
        Anagnostou et al. \citesupp{1anagnostou2020multimodal} & Lung & TRP & Multi-Omic, CR & TCGA & Nat. Can. 2020 \\ \hline
        Lara et al. \citesupp{1lara2020multimodal} & Prostate & Grading, Retrieval & Path, Text & TCGA & MICCAI 2020 \\ \hline
        Mo et al. \citesupp{1mo2020multimodal} & Liver & SEG & Multi-MRI & Private & MICCAI 2020 \\ \hline
        He et al. \citesupp{1he2020feasibility} & Lung & SA & CR, CT & UCI & Info. Fusion 2020 \\ \hline
        Zhou et al. \citesupp{1zhou2020one} & Brain & SEG & Multi-MRI & BraTS & TIP 2020 \\ \hline
        Akil et al. \citesupp{1akil2020fully} & Brain & SEG & Multi-MRI & BraTS & MIA 2020 \\ \hline
        Peng et al. \citesupp{1peng2020multi} & Sarcomas & MD & PET, CT & TCIA & MICCAI 2020  \\ \hline
        He et al. \citesupp{1he2020feasibility} & Prostate, Lung & Toxicity, SA & CR, CT & Private & Inform. Fusion 2020  \\ \hline
        Bi et al. \citesupp{1bi2020multi} & Skin & Subtyping & Multi-image & 7-point Checklist & PR 2020  \\ \hline
        Shao et al. \citesupp{1shao2020multi} & Multiple & Staging, SA & CR, Omic, Path & TCGA & MIA 2020  \\ \hline
        Zheng et al. \citesupp{1zheng2020deep} & Breast & Lymph node status & Multi-US & Private & Nat. Commun. 2020  \\ \hline
        Silva et al. \citesupp{1silva2020pan} & Multiple & SA & CR, Omic, Path & TCGA & ISBI 2020  \\ \hline
        Ibtehaz et al. \citesupp{1ibtehaz2020multiresunet} & Brain & SEG & Multi-MRI & BraTS & Neural Networks 2020 \\ \hline
        Mehrtash et al. \citesupp{1mehrtash2020confidence} & Brain, Prostate & SEG & Multi-MRI & BraTS, PROSTATEx & TMI 2020 \\ \hline
        Zhang et al. \citesupp{1zhang2020exploring} & Brain & SEG & Multi-MRI & BraTS & TIP 2020  \\ \hline
        Naser et al. \citesupp{1naser2020brain} & Brain & SEG, Grading & Multi-MRI & TCIA & CBM 2020 \\ \hline
        Zhang et al. \citesupp{1zhang2021modality} & Brain, Liver & SEG & Multi-MRI, Multi-CT & BraTS, private & MICCAI 2021  \\ \hline
        Yala et al. \citesupp{1yala2021toward} & Breast & Screening & Multi-MMG & Private & Sci. Transl. Med. 2021 \\ \hline
        Han et al. \citesupp{1han2021histologic} & Lung & Subtyping & PET, CT & Private & EJNMMI 2021  \\ \hline
        Arya et al. \citesupp{1arya2021multi} & Breast & SA & CR, Omic & METABRIC, TCGA & KBS 2021  \\ \hline
        Isensee et al. \citesupp{1isensee2021nnu} & Multiple & SEG & Multi-MRI & BraTS & Nature Methods 2021 \\ \hline
        Zhao et al. \citesupp{1zhao2021united} & Liver & SEG, DET & Multi-MRI & Private & MIA 2021  \\ \hline
        Zhang et al. \citesupp{1zhang2021cross1} & Brain & SEG & Multi-MRI & BraTS & PR 2021 \\ \hline
        Cui et al. \citesupp{1cui2020unified} & Brain & SEG & Multi-MRI & MRBrainS & TMI 2021 \\ \hline
        Braman et al. \citesupp{1braman2021deep} & Brain & SA & MRI, Path, Omic, CR & TCGA, TCIA & MICCAI 2021 \\ \hline
        Holste et al. \citesupp{1holste2021end} & Breast & Screening & CR, MRI & Private & ICCV workshops 2021 \\ \hline
        Luo et al. \citesupp{1luo2020hdc} & Brain & SEG & Multi-MRI & BraTS & JBHI 2021 \\ \hline
        Hou et al. \citesupp{1hou2021integration} & Prostate & MD & CR, MRI & Private & EBioMedicine 2021 \\ \hline
        Ning et al. \citesupp{1ning2021multi} & Multiple & SA & Path, Omic & TCGA & TNNLS 2021 \\ \hline
        Bortolini et al. \citesupp{1bortolini2021multimodal} & Breast & SA, MD & Multi-Omic & Private & npj Breast Cancer 2021 \\ \hline
        Rossi et al. \citesupp{1rossi2020multi} & Prostate & Retrieval, Screening & Multi-MRI & Private & TMI 2021 \\ \hline
        Fu et al. \citesupp{1fu2021multimodal} & Lung & SEG & PET, CT & Private & JBHI 2021 \\ \hline
        Chen et al. \citesupp{1chen2021multimodal} & Multiple & SA & Path, Omic & TCGA & ICCV 2021 \\ \hline
        Lv et al. \citesupp{1lv2021pg} & Colorectal & SA & Path, Omic & TCGA & BIBM 2021 \\ \hline
        Jin et al. \citesupp{1jin2021predicting} & Rectal & TRP, SEG & Multi-MRI & Private & Nat. comm. 2021 \\ \hline
        Qian et al. \citesupp{1qian2021prospective} & Breast & Screening & Multi-US & Private & Nat. Bio. Eng. 2021 \\ \hline
        Ding et al. \citesupp{1ding2020scnet} & Gastric & Screening & EI, CR & Private & JBHI 2021 \\ \hline
        Zhang et al. \citesupp{18911262} & Bladder & Grading & Path, CR & BCIDR & TPAMI 2021 \\ \hline
        Wang et al. \citesupp{1wang2021transbts} & Brain & SEG &  Multi-MRI & BraTS & MICCAI 2021 \\ \hline
        Schulte et al. \citesupp{1schulte2021integration} & Multiple & GP & Multi-Omic & TCGA & Nat. Mach. Intell. 2021  \\ \hline
        Chai et al. \citesupp{1chai2021integrating} & Multiple & SA & Multi-Omic & TCGA & CBM 2021  \\ \hline
        Dong et al. \citesupp{1dong2021polarization} & Cervical & Grading & Multi-Path & Private & TMI 2021  \\ \hline
        Zheng et al. \citesupp{1zheng2022automatic} & Liver & SEG & Multi-MRI & Private & TMI 2022  \\ \hline
        Jiang et al. \citesupp{1jiang2022one} & Lung & Registration, SEG & Multi-CT & Private & TMI 2022  \\ \hline
        Li et al. \citesupp{1li2022npcnet} & H\&N & Seg, MD & Multi-MRI & Private & TMI 2022  \\ \hline
        Cheng et al. \citesupp{1cheng2022fully} & Brain & Genotyping, SEG & Multi-MRI & BraTS & TMI 2022 \\ \hline
        Tan et al. \citesupp{1tan2022multi} & Brain & SA, Grading & Path, Omic & TCGA & AI In Medicine 2022 \\ \hline
        Li et al. \citesupp{1li2022adaptive} & Liver & Grading & Multi-MRI & Private & JBHI 2022 \\ \hline
        Zhuang et al. \citesupp{1zhuang2021aprnet} & Brain & SEG & Multi-MRI & MRBrainS & JBHI 2022 \\ \hline
        Qiao et al. \citesupp{1qiao2022breast} & Breast & MD, GP, Grading & MRI-US & Private & JBHI 2022 \\ \hline
        Zhang et al. \citesupp{1zhang2021cross} & Prostate & SEG & Multi-MRI & Private & JBHI 2022 \\ \hline
        Chen et al. \citesupp{1chen2022dual} & Multiple & Subtyping, Screening & Multi-Path & Private & TMI 2022 \\ \hline
        Li et al. \citesupp{1li2022hfbsurv} & Multiple & SA & Path, Omic & TCGA & Bioinformatics 2022 \\ \hline
        Liu et al. \citesupp{1liu2022multimodal} & Brain & SEG & Multi-MRI & BraTS & MICCAI 2022 \\ \hline
        Boehm et al. \citesupp{1boehm2022multimodal} & Ovarian  & SA, TRP & CT, Path, Omic & Private & Nature Cancer 2022 \\ \hline
        Cheng et al. \citesupp{19478224} & Brain & Grading & Multi-MRI & BraTS, Private & JBHI 2022 \\ \hline
        Vanguri et al. \citesupp{1vanguri2022multimodal} & Lung & TRP & CT, Path, Omic & Private & Nature Cancer 2022 \\ \hline
        Han et al. \citesupp{1han2022multimodal} & Multiple & Subtyping, Grading & Multi-Omic & TCGA & CVPR 2022 \\ \hline
        Zheng et al. \citesupp{1zheng2022multi} & H\&N & SA & CR, CT & Private & MICCAI 2022 \\ \hline
        Ding et al. \citesupp{1ding2021mvfusfra} & Brain & SEG & Multi-MRI & BraTS & JBHI 2022 \\ \hline
        Chen et al. \citesupp{1chen2022pan} & Multiple & SA & Path, Omic & TCGA & Cancer cell 2022 \\ \hline
        Chen et al. \citesupp{19186053} & Multiple & SA, Grading & Path, Omic & TCGA & TMI 2022 \\ \hline
        Chen et al. \citesupp{1chen2022scaling} & Multiple & Subtyping, SA & Path & TCGA & CVPR 2022 \\ \hline
        Fang et al. \citesupp{1fang2021self} & Brain & SEG & Multi-MRI & BraTS & JBHI 2022 \\ \hline
        Saeed et al. \citesupp{1saeed2022tmss} & H\&N & SEG, SA & CT, PET, CR & HECKTOR & MICCAI 2022 \\ \hline
        Wang et al. \citesupp{1wang2021interpretability} & Skin & Subtyping & DI, CR & HAM10000 & TCYB 2022 \\ \hline
        Tang et al. \citesupp{1tang2022improving} & Lung & Grading & CT, CR & LUNA16, LIDC-IDRI & Info. Fusion 2022 \\ \hline
        Zhu et al. \citesupp{1zhu2023brain} & Brain & SEG & Multi-MRI & BraTS & Info. Fusion 2022 \\ \hline
        Huang et al. \citesupp{1huang2022evidence} & Brain & SEG & Multi-MRI & BraTS & MICCAI 2022  \\ \hline
        Lehman et al. \citesupp{1lehman2022deep} & Breast & Screening & Multi-MMG & Private & JNCI 2022 \\ \hline
        Yala et al. \citesupp{1yala2022multi} & Breast & Screening & Multi-MMG & Private & J. Clin. Oncol. 2022 \\ \hline
        Luo et al. \citesupp{1luo2022semi} & Brain & SEG & Multi-MRI & BraTS & MIA 2022 \\ \hline
        Sammut et al. \citesupp{1sammut2022multi} & Breast & TRP & Multi-Omic & Private & Nature 2022  \\ \hline
        Pereira et al. \citesupp{1ullah2022cascade} & Brain & SEG & Multi-MRI & BraTS & Inform. Sciences 2022 \\ \hline
        Schulte et al. \citesupp{1solari2022added} & Prostate & Staging & PET, MRI & Private & EJNMMI 2022  \\ \hline
        Xiao et al. \citesupp{1xiao2022task} & Liver & DET & Multi-MRI & Private & MIA 2022  \\ \hline
        Yang et al. \citesupp{1yang2022multi} & Breast & Screening & Multi-MRI & Private & CBM 2022  \\ \hline
        Zhao et al. \citesupp{1zhao2023adaptive} & Multiple & SA & Path, Omic & TCGA & Brief. in Biom. 2023  \\ \hline
        Zhang et al. \citesupp{1zhang2023twist} & Brain, H\&N & SEG & Multi-MRI & Private & CBM 2023  \\ \hline
        Yang et al. \citesupp{1yang2023triple} & Breast & Screening & Multi-MRI & Private & PR 2023  \\ \hline
        Zhu et al. \citesupp{1zhu2023samms} & Multiple & SA & CR, Omic, Path & TCGA & BIBM 2023  \\ \hline
        Meng et al. \citesupp{1meng2023merging} & H\&N & SA, SEG & PET, CT & HECKTOR & MICCAI 2023  \\ \hline
        Xiao et al. \citesupp{1xiao2023edge} & Liver & Seg, Uncertainty & Multi-MRI & private & MICCAI 2023  \\ \hline
        Zhao et al. \citesupp{1zhao2023learning} & Brain, Liver & SEG & Multi-MRI & BraTS, Private & MICCAI 2023  \\ \hline
        Zhou et al. \citesupp{1zhou2023nnformer} & Brain & SEG & Multi-MRI & BraTS & TIP 2023 \\ \hline
        Zhuang et al. \citesupp{1zhuang20223d} & Brain  & SEG & Multi-MRI & BraTS & JBHI 2023 \\ \hline
        Vagenas et al. \citesupp{1vagenas2022decision} & Skin & MD & PET, CT & Private, autoPET & JBHI 2023 \\ \hline
        Hu et al. \citesupp{1hu2023multi} & Lung & Staging, MD & CR, Multi-CT & Private & JBHI 2023 \\ \hline
        Zhou et al. \citesupp{1zhou2023transformer} & lung & Screening & CT, Text & Private & Nat. Bio. Eng. 2023 \\ \hline
        Yue et al. \citesupp{1yue2023adaptive} & Brain & SEG & Multi-MRI & BraTS & JBHI 2023 \\ \hline
        Wu et al. \citesupp{1wu2023camr}  & Multiple & SA & Path, Omic & TCGA & Bioinformatics 2023 \\ \hline
        Lin et al. \citesupp{1lin2023ckd} & Brain & SEG & Multi-MRI & BraTS & TMI 2023 \\ \hline
        Zhou et al. \citesupp{1zhou2023coco} & Lung & SEG & Multi-MRI & Private & JBHI 2023 \\ \hline
        Zhou et al. \citesupp{1zhou2023cross} & Multiple & SA & Path, Omic & TCGA & ICCV 2023 \\ \hline
        Fan et al. \citesupp{1fan2023exploring} & Liver & Staging & Multi-MRI & Private & ACMMM 2023 \\ \hline
        Yang et al. \citesupp{1yang2023flexible} & Brain & SEG & Multi-MRI & BraTS, ISLES & JBHI 2023 \\ \hline
        \multirow{2}{*}{Shi et al. \citesupp{1shi2023h}} & H\&N / & \multirow{2}{*}{SEG} & PET, CT / & HECKTOR / & \multirow{2}{*}{MICCAI 2023} \\ 
         & Prostate &  & Multi-MRI & PICAI &  \\ \hline
        \multirow{2}{*}{Lin et al. \citesupp{1lin2023lesion}} & Colorectal,  & \multirow{2}{*}{SEG} & \multirow{2}{*}{WLI, NBI} & \multirow{2}{*}{Private} & \multirow{2}{*}{TNNLS 2023} \\ 
        & Esophageal &&&& \\\hline
        Liu et al. \citesupp{1liu2023m} & Colorectal & MSI prediction & WSI, CT & Private & MICCAI workshop 2023 \\ \hline
        Hou et al. \citesupp{1hou2023mfd} & Brain & SEG & Multi-MRI & BraTS & JBHI 2023 \\ \hline
        Liu et al. \citesupp{1liu2023mgct} & Multiple & SA & Path, Omic & TCGA & BIBM 2023 \\ \hline
        Marinov et al. \citesupp{1marinov2023mirror} & Multiple & SEG & CT, PET, Multi-MRI & autoPET, MSD & ICCV workshops 2023 \\ \hline
        Xiang et al. \citesupp{1xiang2022modality} & Lung & SEG & PET, CT & Private & JBHI 2023 \\ \hline
        Zheng et al. \citesupp{1zheng2023multi} & Multiple & Subtyping, Grading & Multi-Omic & TCGA & AAAI 2023 \\ \hline
        Podobnik et al. \citesupp{1podobnik2023multimodal} & H\&N & SEG & CT, MRI & HaN-Seg & MICCAI 2023 \\ \hline
        Nabbi et al. \citesupp{1nabbi2023multimodal} & Multiple & SA & Multi-Omic & Private & Nature Cancer 2023 \\ \hline
        Xu et al. \citesupp{1xu2023multimodal} & Multiple & SA & Path, Omic & TCGA & ICCV 2023 \\ \hline
        Gui et al. \citesupp{1gui2023multimodal} & Kidney & RP & CR, Path, Omic & TCGA, Private & Lancet Digit. Health 2023 \\ \hline
        Ding et al. \citesupp{1ding2023pathology} & Colorectal & SA & Path, Omic & TCGA & MICCAI 2023 \\ \hline
        Zhou et al. \citesupp{1zhou2023rfia} & Oesophageal & Staging, Grading & Path, EI & Private & EAAI 2023 \\ \hline
        Gu et al. \citesupp{1gu2023segcofusion} & Brain & SEG & Multi-MRI & BraTS & JBHI 2023 \\ \hline
        Wang et al. \citesupp{110155265} & Thyroid  & MD, SA & Path, CR & TCGA, private & TMI 2023 \\ \hline
        Nakhli et al. \citesupp{1nakhli2023sparse} & Ovarian, Bladder & SA & Multi-Path & Private & CVPR 2023 \\ \hline
        \multirow{2}{*}{Li et al. \citesupp{1li2023survival}} & Brain /  & \multirow{2}{*}{SA} & Path, MRI / & TCGA /  & \multirow{2}{*}{TMI 2023} \\
         & Gastric  &  & Path, CT, CR & Private &  \\ \hline
        Qu et al. \citesupp{1qu2023qnmf} & Breast & Screening & US, Path & BCW, Medminst & Info. Fusion 2023 \\  \hline
        Kim et al. \citesupp{1kim2023heterogeneous} & Breast & Subtyping & MRI, Text & DUKE, CMMD & AAAI 2023 \\ \hline
        Qayyum et al. \citesupp{1qayyum20223d} & H\&N & Seg, SA & PET, CT, CR & HECKTOR & JBHI 2024 \\ \hline
        Tang et al. \citesupp{1tang2024new} & Colorectal & SA, RP & Multi-MRI & Private & JBHI 2024 \\ \hline
        Shi et al. \citesupp{1shi2023novel} & Lung & SA & Path, Omic & TCGA & JBHI 2024 \\ \hline
        Zhang et al. \citesupp{1zhang2024robust} & Brain & Seg, Fusion & Multi-MRI & MRBrainS & AAAI 2024 \\ \hline
        Yan et al. \citesupp{1yan2024combiner} & Prostate & Detection & Multi-MRI & Private & MIA 2024 \\ \hline
        Xiang et al. \citesupp{1xiang2024development} & Ovarian & Screening & CR, US & Private & Nat. Comm. 2024 \\ \hline
        Fang et al. \citesupp{1fang2024dynamic} & Breast & Subtyping & CR, CT & TCGA & WACV 2024 \\ \hline
        \multirow{2}{*}{Alwazzan et al. \citesupp{1alwazzan2024foaa}} & Brain / & Grading / & Path, Omic / & TCGA / & \multirow{2}{*}{ISBI 2024} \\ 
         & Breast & Subtyping & Multi-MMG, CR & CMMD &  \\ \hline
        Li et al. \citesupp{1li2023lvit} & Lung & SEG & CT, Text & Private & TMI 2024 \\ \hline
        Shi et al. \citesupp{1shi2024mif} & Multiple & SA & Path, Omic & TCGA & JBHI 2024 \\ \hline
        Jaume et al. \citesupp{1jaume2023modeling} & Multiple & SA & Path, Omic & TCGA & CVPR 2024 \\ \hline
        Li et al. \citesupp{1li2023msa} & Breast & Grading & MMG, CR & DDSM, INbreast & JBHI 2024 \\ \hline
        Zhang et al. \citesupp{1zhang2023multi} & H\&N / Brain & SEG & CT,PET / Multi-MRI & HECHTOR / BraTS & TMI 2024 \\ \hline
        Liu et al. \citesupp{1liu2024multimodal} & Brain & SEG & Multi-MRI & BraTS, Private & TIP 2024 \\ \hline
        Meng et al. \citesupp{1meng2024nama} & Prostate & SEG & Multi-MRI & PICAI, Private & AAAI 2024 \\ \hline
        Zhang et al. \citesupp{1zhang2024prototypical} & Multiple & SA & Path, Omic & TCGA & ICLR 2024 \\ \hline
        Donnelly et al. \citesupp{1donnelly2024asymmirai} & Breast & Screening & Multi-MMG & Private & Radiology 2024 \\ \hline
\end{supertabular}
\end{center}

\clearpage
\twocolumn

\begin{table*}[!ht]
    \centering
    \setlength\tabcolsep{1pt}
    \caption{Summary of reviewed representative papers on integrating incomplete multimodal data. SA: Survival analysis; Seg: Segmentation; Path: Pathology; Omic: Omics Data; CR: Clinical Records; WNR: Wisconsin Neurodevelopment Rhesus dataset.}
    \label{tab:missing}
    \begin{tabular}{|l|l|l|l|l|l|l|}
    \hline
        \textbf{Approach} & \textbf{Taxonomy} & \textbf{Cancer type} & \textbf{Application} & \textbf{Modality} & \textbf{Dataset} & \textbf{Publication}  \\ \hline
        Havaei et al. \citesupp{1havaei2016hemis} & Robustness enhancement
        & Brain & SEG & Multi-MRI & BraTS & MICCAI 2016 \\ \hline
        Chartsias et al. \citesupp{1chartsias2017multimodal} & Data generation& Brain & SEG & Multi-MRI & ISLES, BraTS  & TMI 2017 \\ \hline
        Van et al. \citesupp{1van2018learning} & Multi-task learning& Brain & SEG & Multi-MRI & BraTS  & TMI 2018 \\ \hline
        Sharma et al. \citesupp{1sharma2019missing} & Data generation& Brain & SEG & Multi-MRI & ISLES, BraTS  & TMI 2019 \\ \hline
        Dorent et al. \citesupp{1dorent2019hetero} & Multi-task learning& Brain & SEG & Multi-MRI & BraTS  & MICCAI 2019 \\ \hline
        Chen et al. \citesupp{1chen2019robust} & Multi-task learning & Brain & SEG & Multi-MRI & BraTS & MICCAI 2019 \\ \hline
        Shen et al. \citesupp{1shen2019brain} & Knowledge distillation & Brain & SEG & Multi-MRI & BraTS & IPMI 2019 \\ \hline
        Zhou et al. \citesupp{1zhou2020hi} & Data generation & Brain & SEG & Multi-MRI & BraTS & TMI 2020 \\ \hline
        Vilardell et al. \citesupp{1vilardell2020missing} & Data generation & Breast & SA & CR & Private & AI in Medicine 2020 \\ \hline
        Zhou et al. \citesupp{1zhou2020brain} & Multi-task learning & Brain & SEG & Multi-MRI & BraTS & MICCAI 2020 \\ \hline
        Hu et al. \citesupp{1hu2020knowledge} & Knowledge distillation & Brain & SEG & Multi-MRI & BraTS & MICCAI 2020 \\ \hline
        Shen et al. \citesupp{1shen2020multi} & Data generation & Brain & SEG & Multi-MRI & BraTS, PROSTATEx & TMI 2021 \\ \hline
        Hamghalam et al. \citesupp{1hamghalam2021modality} & Data generation & Brain & SEG & Multi-MRI & BraTS & MICCAI 2021 \\ \hline
        Gao et al. \citesupp{1gao2021lung} & Data generation & Lung & Screening & CT, CR & NLST, Private & MICCAI 2021 \\ \hline
        Peng et al. \citesupp{1peng2021multi} & Data generation & Brain & SEG & Multi-MRI & BraTS & JBHI 2021 \\ \hline
        Huang et al. \citesupp{1huang2021aw3m} & Feature generation & Brain & SEG & Multi-MRI & Private & MIA 2021 \\ \hline
        Nikhilanand et al. \citesupp{1arya2021generative} & Feature generation & Breast & SA & CR, Omic & METABRIC, TCGA & TCBB 2021 \\ \hline
        Zhou et al. \citesupp{1zhou2021latent} & Multi-task learning & Brain & SEG & Multi-MRI & BraTS & TIP 2021 \\ \hline
        Rahimpour et al. \citesupp{1rahimpour2021cross} & Knowledge distillation & Brain & SEG & Multi-MRI & Private, BraTS & TBME 2021 \\ \hline
        Wang et al. \citesupp{1wang2021acn} & Knowledge distillation & Brain & SEG & Multi-MRI & BraTS & MICCAI 2021 \\ \hline
        Ning et al. \citesupp{1ning2021relation} & Robustness enhancement & Multiple & SA & Path, Omic & TCGA & TMI 2021 \\ \hline
        Vadacchino et al. \citesupp{1vadacchino2021had} & Knowledge distillation & Brain & SEG & Multi-MRI & BraTS & MIDL 2021 \\ \hline
        Ding et al. \citesupp{1ding2021rfnet} & Robustness enhancement & Brain & SEG & Multi-MRI & BraTS & ICCV 2021 \\ \hline
        Vale et al. \citesupp{1vale2021long} & Robustness enhancement & Multiple & SA & Path, Omic, CR & TCGA & Sci. Rep. 2021 \\ \hline
        Zhou et al. \citesupp{1zhou2022missing} & Data generation & Brain & SEG & Multi-MRI & BraTS & PRL 2022 \\ \hline        
        Dalmaz et al. \citesupp{1dalmaz2022resvit} & Data generation & Brain & SEG & Multi-MRI & BraTS, IXI & TMI 2022 \\ \hline
        Zhou et al. \citesupp{1biology11030360} & Feature generation & Brain & SEG & Multi-MRI & BraTS & Biology 2022 \\ \hline
        Jeong et al. \citesupp{1jeong2022region} & Multi-task learning & Brain & SEG & Multi-MRI & BraTS, ISLES & ACCV 2022 \\ \hline
        Cui et al. \citesupp{1cui2022survival} & Multi-task learning & Multiple & SA & Path, Omic, MRI, CR & TCGA & MICCAI 2022 \\ \hline
        Azad et al. \citesupp{1azad2022smu} & Knowledge distillation & Brain & SEG & Multi-MRI & BraTS & MIDL 2022 \\ \hline
        Yang et al. \citesupp{1yang2022d} & Knowledge distillation & Brain & SEG & Multi-MRI & BraTS & TMI 2022 \\ \hline
        Ning et al. \citesupp{1ning2022mutual} & Knowledge distillation & Multiple & SA & Path, Omic & TCGA & TPAMI 2022 \\ \hline
        Liu et al. \citesupp{1liu2022moddrop} & Robustness enhancement & Brain & SEG & Multi-MRI & UMCL & MICCAI 2022 \\ \hline
        Zhang et al. \citesupp{1zhang2022mmformer} & Robustness enhancement & Brain & SEG & Multi-MRI & BraTS & MICCAI 2022 \\ \hline
        Zhao et al. \citesupp{1zhao2022modality} & Robustness enhancement & Brain & SEG & Multi-MRI & BraTS & MICCAI 2022 \\ \hline
        Yang et al. \citesupp{1yang2023learning} & Data generation & Brain & SEG & Multi-MRI & BraTS & TMI 2023 \\ \hline
        Wu et al. \citesupp{1wu2023collaborative} & Data generation & Brain & SEG & Multi-MRI & WNR & MICCAI 2023 \\ \hline
        Yuan et al. \citesupp{1yuan2023rethinking} & Data generation & Brain & SEG & Multi-MRI & BraTS & MICCAI 2023 \\ \hline
        Yael et al. \citesupp{1moshe2023handling} & Data generation & Brain & Grading & Multi-MRI & Private, BraTS & JMRI 2023 \\ \hline
        Diao et al. \citesupp{1diao2023joint} & Feature generation & Brain & SEG & Multi-MRI & BraTS & CBM 2023 \\ \hline
        Wang et al. \citesupp{1wang2023multi} & Feature generation & Brain & SEG & Multi-MRI & BraTS & CVPR 2023 \\ \hline
        Hou et al. \citesupp{1hou2023hybrid} & Feature generation & Multiple & SA & Path, Omic, CR & TCGA & TMI 2023 \\ \hline
        Liu et al. \citesupp{1liu2023learning} & Multi-task learning & Brain & SEG & Multi-MRI & BraTS & CBM 2023 \\ \hline
        Zhou et al. \citesupp{1zhou2023feature} & Multi-task learning & Brain & SEG & Multi-MRI & BraTS & PR 2023 \\ \hline
        Liu et al. \citesupp{1liu2023m3ae} & Multi-task learning & Brain & SEG & Multi-MRI & BraTS & AAAI 2023 \\ \hline
        Choi et al. \citesupp{1choi2023single} & Knowledge distillation & Brain & SEG & Multi-MRI & BraTS & Comput. Meth. Prog. Bio. 2023 \\ \hline
        Wang et al. \citesupp{1wang2023prototype} & Knowledge distillation & Brain & SEG & Multi-MRI & BraTS & ICASSP 2023 \\ \hline
        Konwer et al. \citesupp{1konwer2023enhancing} & Knowledge distillation & Brain & SEG & Multi-MRI & BraTS & ICCV 2023 \\ \hline
        Zhang et al. \citesupp{1zhang2023multi} & Knowledge distillation & Brain & SEG & Multi-MRI & BraTS & TMI 2023 \\ \hline
        Liu et al. \citesupp{1liu2023one} & Data generation & Brain & SEG & Multi-MRI & IXI, BraTS & TMI 2023 \\ \hline
        Wang et al. \citesupp{1wang2023learnable} & Knowledge distillation & Brain & SEG & Multi-MRI & BraTS & MICCAI 2023 \\ \hline
        Qiu et al. \citesupp{1qiu2023scratch} & Knowledge distillation & Brain & SEG & Multi-MRI & BraTS & ICCV 2023 \\ \hline
        Feng et al. \citesupp{1feng2023brain} & Robustness enhancement & Brain & SEG & Multi-MRI & BraTS & J. Digit. Imaging 2023 \\ \hline
        Qiu et al. \citesupp{1qiu2023modal} & Robustness enhancement & Brain & SEG & Multi-MRI & BraTS & MM 2023 \\ \hline
        Liu et al. \citesupp{1liu2023sfusion} & Robustness enhancement & Brain & SEG & Multi-MRI & BraTS & MICCAI 2023 \\ \hline
        Shi et al. \citesupp{1shi2023m} & Robustness enhancement & Brain & SEG & Multi-MRI & BraTS & JBHI 2023 \\ \hline
        Meng et al. \citesupp{1meng2024multi} & Data generation & Brain & SEG & Multi-MRI & BraTS, IXI & TMI 2024 \\ \hline
        Ting et al. \citesupp{1ting2023multimodal} & Feature generation & Brain & SEG & Multi-MRI & BraTS & JBHI 2024 \\ \hline
        Qiu et al. \citesupp{1qiu2024dual} & Feature generation & Brain & Grading, SA & Path, Omic & TCGA & Phys. Med. Biol. 2024 \\ \hline
        Wang et al. \citesupp{1wang2024mgiml} & Sample retrieval & Multiple & Grading & Path, CT & TCGA & TMI 2024 \\ \hline
        Karimijafarbigloo et al. \citesupp{1karimijafarbigloo2024mmcformer} & Knowledge distillation & Brain & SEG & Multi-MRI & BraTS & MIDL 2024 \\ \hline
        Zhang et al. \citesupp{1zhang2024mhd} & Knowledge distillation & Lung & Grading & Path, CT & CPTAC & JBHI 2024 \\ \hline
        Zhang et al. \citesupp{1zhang2024scalable} & Knowledge distillation & Brain & SEG & Multi-MRI & BraTS & Artif. Intell. Med. 2024 \\ \hline
        Jagadeesh et al. \citesupp{1jagadeesh2024brain} & Robustness enhancement & Brain & SEG & Multi-MRI & BraTS & Appl. Soft Comput.  2024 \\ \hline
        Xing et al. \citesupp{1xing2024pre} & Robustness enhancement & Brain & SEG & Multi-MRI & BraTS & ICASSP  2024 \\ \hline
        Liu et al. \citesupp{1liu2024mixture} & Robustness enhancement & Brain & SEG & Multi-MRI & BraTS & MBEC 2024 \\ \hline
        Qiu et al. \citesupp{1qiu2024mmmvit} & Robustness enhancement & Brain & SEG & Multi-MRI & BraTS & Biomed. Signal Proces. 2024 \\ \hline
        Zhang et al. \citesupp{1zhang2024tmformer} & Robustness enhancement & Brain & SEG & Multi-MRI & BraTS & AAAI 2024 \\ \hline
        
    \end{tabular}
\end{table*}

\begin{table*}[!ht]
    \centering
    \setlength\tabcolsep{25pt}
    \caption{Summary of reviewed survey papers related to multimodal data integration on precision oncology.}
    \label{tab:survey}
    \begin{tabular}{|l|l|l|l|l|l|l|}
    \hline
        \textbf{Survey}   & \textbf{Journal} & \textbf{Year}  \\ \hline
        Sarris et al. \citesupp{1sarris2014multimodal} &  Cancer Treatment Reviews & 2014 \\ \hline
        Yankeelov et al. \citesupp{1yankeelov2014quantitative} &  Nature Reviews Clinical Oncology & 2014 \\ \hline
        Sauter et al. \citesupp{1sauter2015image} &  European Journal of Nuclear Medicine and Molecular Imaging & 2015 \\ \hline
        Lahat et al. \citesupp{1lahat2015multimodal} &  Proceedings of the IEEE & 2015 \\ \hline
        Ravi et al. \citesupp{1ravi2016deep} &  Journal of Biomedical and Health Informatics & 2016 \\ \hline
        Madabhushi et al. \citesupp{1madabhushi2016image} &  Medical Image Analysis & 2016 \\ \hline
        Bibault et al. \citesupp{1bibault2016big} &  Cancer Letters & 2016 \\ \hline
        Ramachandram et al. \citesupp{1ramachandram2017deep} &  IEEE Signal Processing Magazine & 2017 \\ \hline
        Bejnordi et al. \citesupp{1bejnordi2017diagnostic} &  Jama & 2017 \\ \hline
        Dora et al. \citesupp{1dora2017state} &  IEEE Reviews in Biomedical Engineering & 2017 \\ \hline
        Baltruvsaitis et al. \citesupp{1baltruvsaitis2018multimodal} &  IEEE Transactions on Pattern Analysis and Machine Intelligence & 2018 \\ \hline
        Hu et al. \citesupp{1hu2018deep} &  Pattern Recognition & 2018 \\ \hline
        Hamidinekoo et al. \citesupp{1hamidinekoo2018deep} &  Medical Image Analysis & 2018 \\ \hline
        Ghaffari et al. \citesupp{1ghaffari2019automated} &  IEEE Reviews in Biomedical Engineering & 2019 \\ \hline
        Ghafoor et al. \citesupp{1ghafoor2019multimodality} &  Journal of Nuclear Medicine & 2019 \\ \hline
        Bi et al. \citesupp{1bi2019artificial} &  CA: A Cancer Journal for Clinicians & 2019 \\ \hline
        Zhang et al. \citesupp{1zhang2020advances} &  Information Fusion & 2020 \\ \hline
        Panayides et al. \citesupp{1panayides2020ai} &  Journal of Biomedical and Health Informatics & 2020 \\ \hline
        Huang et al. \citesupp{1huang2020fusion} &  npj Digital Medicine & 2020 \\ \hline
        Boehm et al. \citesupp{1boehm2022harnessing} &  Nature Reviews Cancer & 2021 \\ \hline
        Muhammad et al. \citesupp{1muhammad2021comprehensive} &  Information Fusion & 2021 \\ \hline
        Muhammad et al. \citesupp{1muhammad2020deep} &  IEEE Transactions on Neural Networks and Learning Systems & 2021 \\ \hline
        Kang et al. \citesupp{1kang2022roadmap} &  Briefings in Bioinformatics & 2022 \\ \hline
        Lipkova et al. \citesupp{1lipkova2022artificial} &  Cancer cell & 2022 \\ \hline
        Behrad et al. \citesupp{1behrad2022overview} &  Expert Systems with Applications & 2022 \\ \hline
        Stahlschmidt et al. \citesupp{1stahlschmidt2022multimodal} &  Briefings in Bioinformatics & 2022 \\ \hline
        Kline et al. \citesupp{1kline2022multimodal} &  npj Digital Medicine & 2022 \\ \hline
        Soenksen et al. \citesupp{1soenksen2022integrated} &  npj Digital Medicine & 2022 \\ \hline
        Acosta et al. \citesupp{1acosta2022multimodal} &  Nature Medicine & 2022 \\ \hline
        Tong et al. \citesupp{1tong2023integrating} &  IEEE Reviews in Biomedical Engineering  & 2023 \\ \hline
        Tong et al. \citesupp{1eisenmann2023winner} &  IEEE International Conference on Computer Vision and Pattern Recognition & 2023 \\ \hline
        Ranjbarzadeh et al. \citesupp{1ranjbarzadeh2023brain} &  Computers in Biology and Medicine & 2023 \\ \hline
        Qiu et al. \citesupp{1qiu2023large} & Journal of Biomedical and Health Informatics & 2023 \\ \hline
        Steyaert et al. \citesupp{1steyaert2023multimodal} &  Nature Machine Intelligence & 2023 \\ \hline
        Shamshad et al. \citesupp{1shamshad2023transformers} &  Medical Image Analysis & 2023 \\ \hline
        Azad et al. \citesupp{1azad2023advances} &  Medical Image Analysis & 2023 \\ \hline
        Mazurowski et al. \citesupp{1mazurowski2023segment} &  Medical Image Analysis & 2023 \\ \hline
        Shaik et al. \citesupp{1shaik2023survey} &  Information Fusion & 2023 \\ \hline
        Wu et al. \citesupp{1wu2024big} &  Trends in cancer & 2024 \\ \hline
        Luo et al. \citesupp{1luo2024deep} &  IEEE Reviews in Biomedical Engineering & 2024 \\ \hline
        Zhao et al. \citesupp{1zhao2024review} &  Information Fusion & 2024 \\ \hline
        Lotter et al. \citesupp{1lotter2024artificial} &  Cancer Discovery & 2024 \\ \hline
    \end{tabular}
\end{table*}

\end{document}